\documentclass[preprint,12pt,authoryear]{elsarticle}

\usepackage[english]{babel}
\usepackage[ruled,vlined]{algorithm2e}

\usepackage[letterpaper,top=2cm,bottom=2cm,left=3cm,right=3cm,marginparwidth=1.75cm]{geometry}

\usepackage[authoryear,round]{natbib}
\usepackage{amsmath}
\usepackage{amssymb}
\usepackage{multirow}
\usepackage{graphicx}
\usepackage{ulem}
\usepackage{booktabs}
\usepackage[colorlinks=true, allcolors=blue]{hyperref}

\hypersetup{
    colorlinks=true,  % 开启颜色链接
    linkcolor=blue,   % 目录、章节等内部链接颜色
    citecolor=blue,   % 文献引用的链接颜色
    filecolor=blue,   % 文件链接颜色
    urlcolor=blue     % 网址链接颜色
}
\usepackage{color,xcolor}
\definecolor{lightblue}{RGB}{0,139,216}

\usepackage{cleveref}
\Crefname{algorithm}{Algorithm}{Algorithms}

\journal{Medical Image Analysis}

\begin{document}

\title{$U^{2}AD$: Uncertainty-based Unsupervised Anomaly Detection Framework for Detecting T2 Hyperintensity in MRI Spinal Cord}

\author[inst1]{Qi Zhang\fnref{equal}}
\author[inst2]{Xiuyuan Chen\fnref{equal}}
\author[inst3]{Ziyi He\fnref{equal}}
\author[inst2]{Kun Wang\fnref{equal}}
\author[inst4]{Lianming Wu}
\author[inst2]{Hongxing Shen\corref{cor1}}
\author[inst1,inst5,inst6]{Jianqi Sun\corref{cor1}}

\cortext[cor1]{Corresponding authors: Hongxing Shen (email: shenhongxing@renji.com) and Jianqi Sun (email: milesun@sjtu.edu.cn)}

\fntext[equal]{Qi Zhang (email: zhi-bai-shou-hei@sjtu.edu.cn), Xiuyuan Chen (email: chenxiuyuan@renji.com), Ziyi He (hzy12950@rjh.com.cn), and Kun Wang (email: wk52693158@126.com) contributed equally to this work and should be regarded as co-first authors.}

\affiliation[inst1]{organization={School of Biomedical Engineering, Shanghai Jiao Tong University}, city={Shanghai}, country={China}}
\affiliation[inst2]{organization={Department of Spine Surgery, Renji Hospital, School of Medicine, Shanghai Jiao Tong University}, city={Shanghai}, country={China}}
\affiliation[inst3]{organization={Department of Radiology, Ruijin Hospital, Shanghai Jiao Tong University School of Medicine}, city={Shanghai}, country={China}}
\affiliation[inst4]{organization={Department of Radiology, Renji Hospital, School of Medicine, Shanghai Jiao Tong University}, city={Shanghai}, country={China}}
\affiliation[inst5]{organization={National Engineering Research Center of Advanced Magnetic Resonance Technologies for Diagnosis and Therapy (NERC-AMRT)}, city={Shanghai}, country={China}}
\affiliation[inst6]{organization={Med-X Research Institute, Shanghai Jiao Tong University}, city={Shanghai}, country={China}}

\begin{abstract}  
T2 hyperintensities in spinal cord MR images are crucial biomarkers for conditions such as degenerative cervical myelopathy. However, current clinical diagnoses primarily rely on manual evaluation. Deep learning methods have shown promise in lesion detection, but most supervised approaches are heavily dependent on large, annotated datasets. Unsupervised anomaly detection (UAD) offers a compelling alternative by eliminating the need for abnormal data annotations. However, existing UAD methods rely on curated normal datasets and their performance frequently deteriorates when applied to clinical datasets due to domain shifts. We propose an Uncertainty-based Unsupervised Anomaly Detection framework, termed $U^{2}AD$, to address these limitations. Unlike traditional methods, U2AD is designed to be trained and tested within the same clinical dataset, following a “mask-and-reconstruction” paradigm built on a Vision Transformer-based architecture. We introduce an uncertainty-guided masking strategy to resolve task conflicts between normal reconstruction and anomaly detection to achieve an optimal balance. Specifically, we employ a Monte-Carlo sampling technique to estimate reconstruction uncertainty mappings during training. By iteratively optimizing reconstruction training under the guidance of both epistemic and aleatoric uncertainty, U2AD reduces overall reconstruction variance while emphasizing regions. Experimental results demonstrate that U2AD outperforms existing supervised and unsupervised methods in patient-level identification and segment-level localization tasks. This framework establishes a new benchmark for incorporating uncertainty guidance into UAD, highlighting its clinical utility in addressing domain shifts and task conflicts in medical image anomaly detection. Our code is available: \url{https://github.com/zhibaishouheilab/U2AD}.
\end{abstract}


\date{} % 移除日期
\maketitle

\section{Introduction}
%T2 hyperintensity regions in spinal cord MRI images, often associated with cervical spondylopathy, are critical indicators of potential pathology. However, these regions are typically small, blurred, and difficult to distinguish from surrounding tissue, making manual inspection prone to misdiagnosis or missed diagnosis, especially among less experienced clinicians.
As an important part of the central nervous system, the integrity of the structure and function of the cervical spinal cord is crucial for maintaining normal whole-body movements, sensations and various physiological activities in the human body \citep{ahuja2017traumatic}. Magnetic Resonance Imaging (MRI) technology has unique advantages in the diagnosis of cervical spinal cord diseases. Among them, the hyperintensity on T2-weighted images are often one of the important radiological manifestations indicating the presence of lesions such as edema, degeneration, inflammation and tumors in the cervical spinal cord \citep{badhiwala2020degenerative,malhotra2017utility,zivadinov2020cervical,baleriaux1999spinal}. Timely and accurate detection of T2 hyperintensity can facilitate diagnosis and reasonable treatment plans of cervical spinal cord diseases, and improve prognosis of patients ultimately. However, these lesions are typically small, blurred, and difficult to distinguish from surrounding tissue, making manual inspection prone to misdiagnosis or missed diagnosis, especially among less experienced clinicians. Therefore, accurate and automatic detection methods for T2 hyperintensity are highly needed.

\begin{figure*}[!t]
\centerline{\includegraphics[width=\columnwidth]{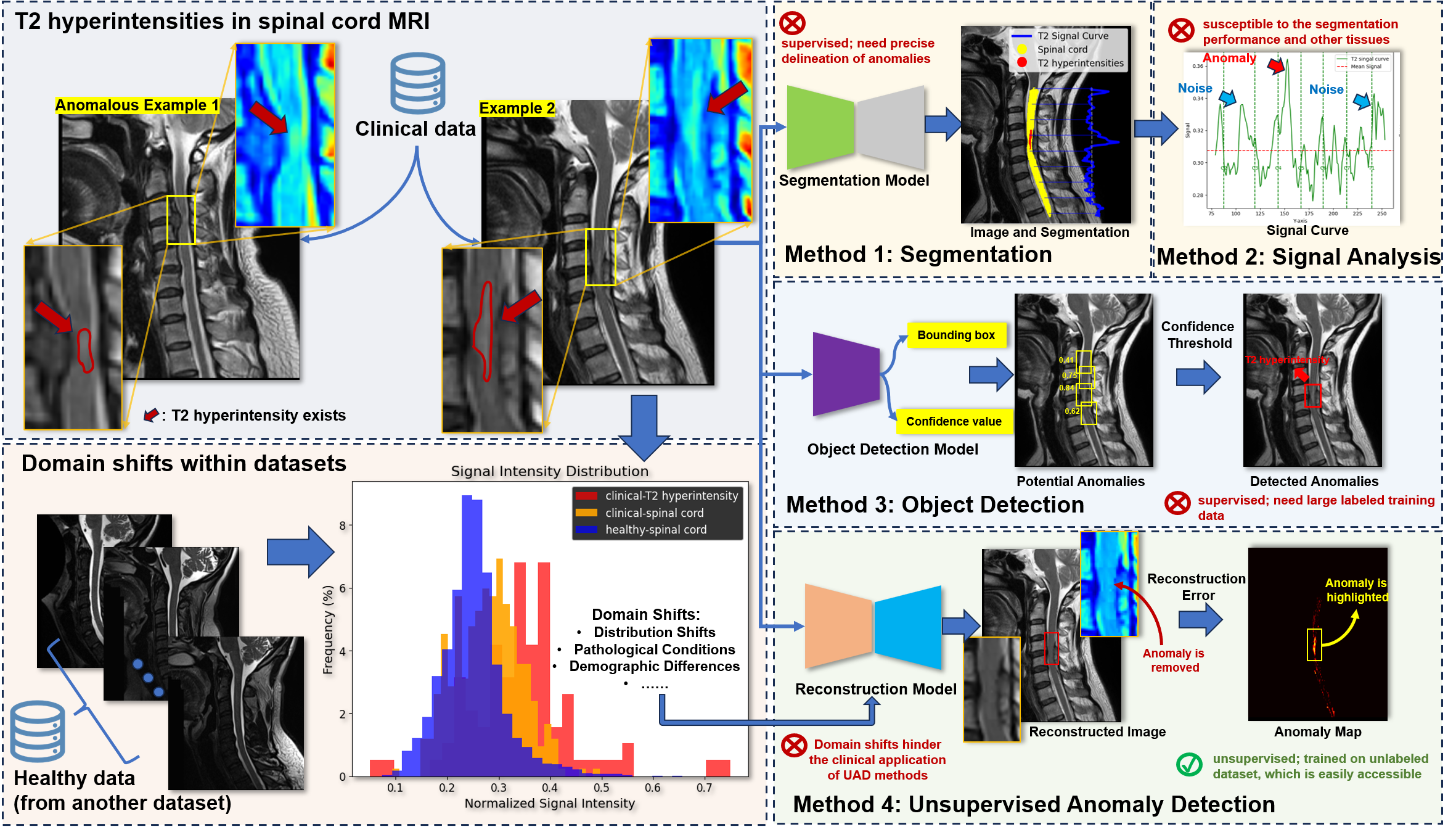}}
\caption{Illustration of T2 hyperintensities in spinal cord MR images, the presence of domain shifts between healthy and clinical datasets, and an overview of different methods used for detecting T2 hyperintensities. The top-left panel highlights anomalous examples with T2 hyperintensities, marked by red outlines. The bottom-left panel depicts the domain shift in signal intensity distribution between healthy datasets and clinical datasets, emphasizing the challenges posed by demographic and pathological differences. The right panel outlines four approaches for T2 hyperintensity detection: (1) segmentation models that require precise anomaly annotations, (2) signal analysis methods prone to noise interference, (3) object detection models reliant on large labeled datasets, and (4) reconstruction-based anomaly detection models, which are unsupervised but can struggle with domain shifts.}
\label{introduction}
\end{figure*}

Deep learning techniques have demonstrated remarkable success in detecting lesions and abnormal regions in medical images, primarily through supervised learning approaches such as segmentation and object detection models (\Cref{introduction}). Models such as U-Net and its variants are extensively used for organ and lesion segmentation but require large annotated datasets and precise manual delineation, which is both time-consuming and labor-intensive. Object detection models like YOLO reduce the need for pixel-level annotations by using coarse bounding boxes for lesion localization. However, their effectiveness depends heavily on extensive labeled datasets and is often limited by variations in lesion contrast, size, and shape. For T2 hyperintensity detection, the scarcity of annotated cases presents additional challenges. Prior studies have explored signal analysis methods \citep{WOS:001017742400001}, but this method is highly sensitive to threshold settings and prone to false positives due to noise from surrounding tissues like cerebrospinal fluid.

Unsupervised anomaly detection (UAD) offers a promising alternative for identifying anomalous regions in medical images by framing the task as one-class classification (OCC) using only normal data for training. These models learn healthy anatomical representations through unsupervised or self-supervised learning and detect anomalies as out-of-distribution deviations in test data. Reconstruction-based UAD methods, such as those employing autoencoders (AEs) and generative adversarial networks (GANs), identify anomalies in areas with high reconstruction errors. While UAD methods have shown success in brain anomaly detection tasks, particularly with AE- and GAN-based approaches \citep{baur2021autoencoders, kascenas2023role}, their application to other tasks, such as detecting T2 hyperintensities in the spinal cord, remains underexplored.

Despite their potential, the conventional UAD paradigm faces two key challenges. First, it relies on curated datasets with exclusively healthy training data and anomalous test data, which is often impractical in real-world clinical settings due to dataset overlap and domain shifts between training and test distributions. Second, UAD models face task conflicts between accurate reconstruction of normal regions and amplifying errors in anomalous regions for detection. Overtraining can mask abnormalities, leading to false negatives, while undertraining may result in false positives from poor reconstruction quality, compromising overall detection performance.

%Despite promising outcomes in anomaly detection, UAD faces two major challenges for clinical application. First, domain shifts hinder model generalization when training and test data distributions differ, as shown in \Cref{introduction}. Second, many UAD methods are pseudo-unsupervised, relying on manually curated healthy training data and anomaly-potential test data, a setup that is often impractical in clinical practice.

In this study, we propose an innovative UAD framework, termed \textbf{Uncertainty-based Unsupervised Anomaly Detection ($U^{2}AD$)}, for detecting T2 hyperintensities in the spinal cord, addressing key challenges in anomaly detection. Unlike conventional methods requiring curated datasets, $U^{2}AD$ is trained and tested on the same clinical dataset using the Vision Transformer (ViT)-based network in a ``mask-and-reconstruction'' paradigm. The framework introduces uncertainty-guided masking strategies to optimize reconstruction training by leveraging aleatoric uncertainty (AU) to amplify errors in anomaly candidates and epistemic uncertainty (EU) to enhance model robustness in low-confidence regions. This dual uncertainty guidance effectively balances normal reconstruction and anomaly sensitivity, improving detection performance. To our knowledge, $U^{2}AD$ is the first UAD framework to integrate both AU and EU, offering significant clinical applicability and advancing T2 hyperintensity detection in spinal cord MRI.

\section{Related Works}

\subsection{Unsupervised Anomaly Detection for Medical Images}

Unsupervised anomaly detection offers an effective alternative to supervised methods, particularly for small clinical datasets, by modeling the distribution of normal anatomy and identifying anomalies as deviations. Generative models like autoencoders (AEs) \citep{shvetsova2021anomaly,baur2021autoencoders,cai2024rethinking}, generative adversarial networks (GANs) \citep{baur2019deep,schlegl2019f}, and denoising diffusion probabilistic models (DDPMs) \citep{behrendt2024leveraging,wyatt2022anoddpm,kascenas2023role} have been widely applied. AE-based methods, leveraging encoder-decoder frameworks, are particularly common but face challenges like overfitting and suboptimal reconstruction of complex regions \citep{liang2024itermask}. Variants such as denoising AEs (DAEs) \citep{kascenas2023role}, variational AEs (VAEs) \citep{kingma2013auto}, and masked AEs (MAEs) \citep{he2022masked} address these issues with strategies like “corrupt-and-reconstruct” or “mask-and-reconstruct”. While UAD has demonstrated success in brain MRI and other imaging modalities, its application to spinal cord anomalies, such as T2 hyperintensities, remains underexplored.

% 把MAE单独拿出来说，MIM其实是一个领域的工作，有很多利用MIM来进行异常检测的，不过没有之间使用MAE这个词而已
\subsection{Mask Image Modeling in Medical Image Analysis}

Masked Image Modeling (MIM), introduced as a self-supervised learning paradigm, has proven effective in medical image analysis tasks such as anomaly detection, segmentation, and classification. By masking image regions and reconstructing the missing parts, MIM enables models to learn robust feature representations. Recent methods like MAEDiff \citep{xu2024maediff} and MOODv2 \citep{li2024moodv2} have shown promise in improving anomaly detection and addressing domain gaps. However, most existing MIM methods rely on random masking strategies, which fail to account for the varying complexities of different image regions. Normal regions require more effective masking in complex areas to enhance learning, while anomalies need reduced masking to amplify reconstruction errors. To overcome these limitations, we propose Uncertainty-Guided Masking Strategies, dynamically adjusting masking to optimize reconstruction fidelity for normal regions and sensitivity for anomalies, thereby advancing MIM for medical image analysis

\subsection{Uncertainty Estimation for Medical Anomaly Detection}

Uncertainty estimation has become an essential tool for improving the reliability and interpretability in anomaly detection. Aleatoric uncertainty reflects data-related noise or ambiguity, such as image quality variations, and highlights potential anomalies during reconstruction. Epistemic uncertainty captures the model's confidence and can guide adaptation to unseen data. In anomaly detection, reconstruction error or variance often serves as a proxy for uncertainty, highlighting inconsistencies in the data and model \citep{seebock2019exploiting,gu2024revisiting}. However, most studies focus on a single type of uncertainty. This work integrates both aleatoric and epistemic uncertainty using pixel-level uncertainty maps derived via Monte Carlo (MC) sampling, enabling iterative optimization of reconstruction training. This dual-guidance approach enhances anomaly detection by balancing reconstruction fidelity and sensitivity.

\section{Methods}

\subsection{Overview}

\begin{figure*}[!t]
\centerline{\includegraphics[width=\columnwidth]{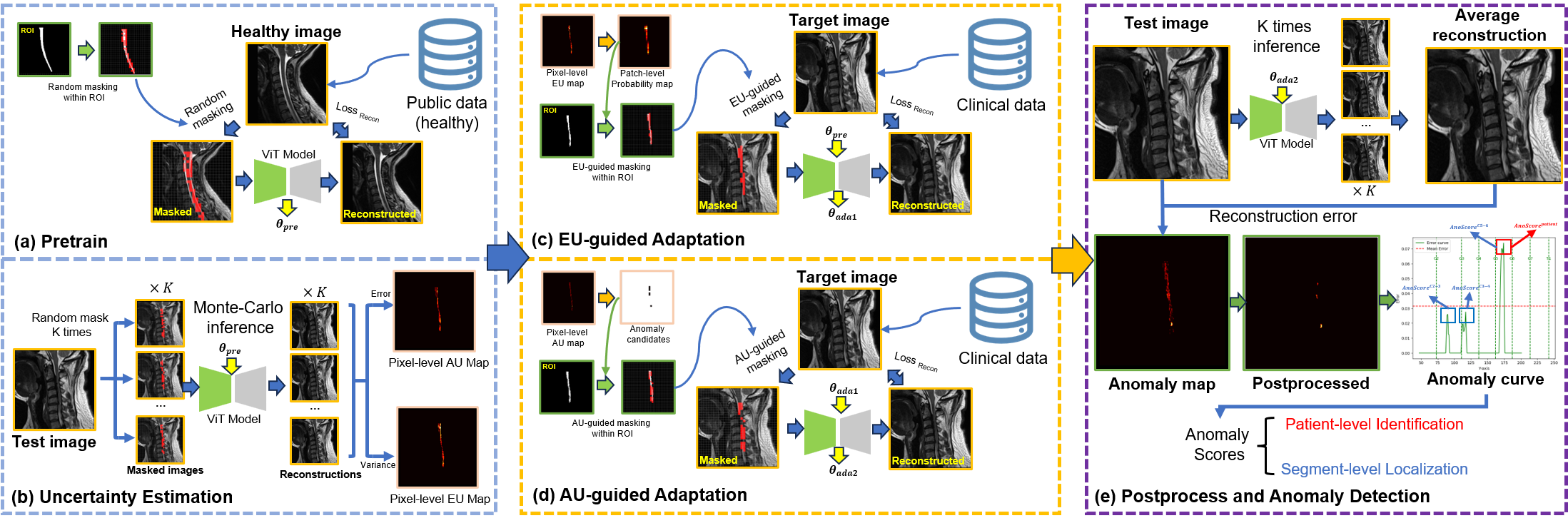}}
\caption{Workflow of the proposed $U^{2}AD$ framework. The model is pretrained on a large-scale healthy dataset, followed by a two-stage adaptation on clinical datasets. The Monte Carlo-based uncertainty estimation is integrated into the mask-and-reconstruct process. Epistemic uncertainty and aleatoric uncertainty guides masking and reconstruction training in the adaptation training. The outputs include patient-level anomaly identification and segment-level localization, with quantitative measures of anomaly scores to assess performance.}
\label{workflow}
\end{figure*}

$U^{2}AD$ adopts an ViT-based asymmetric encoder-decoder architecture, described in \Cref{Network Architecture}.  The workflow, as depicted in \Cref{workflow}, consists of three key phases: (1) \textbf{Pretraining and Uncertainty Estimation}, detailed in \Cref{pretraining}; (2) \textbf{Adaptation Training}, detailed in \Cref{adaptation training}; (3) \textbf{Anomaly detection and localization}, outlined in \Cref{Postprocessing and Anomaly Detection}.

\begin{figure*}[t]
\centerline{\includegraphics[width=\columnwidth]{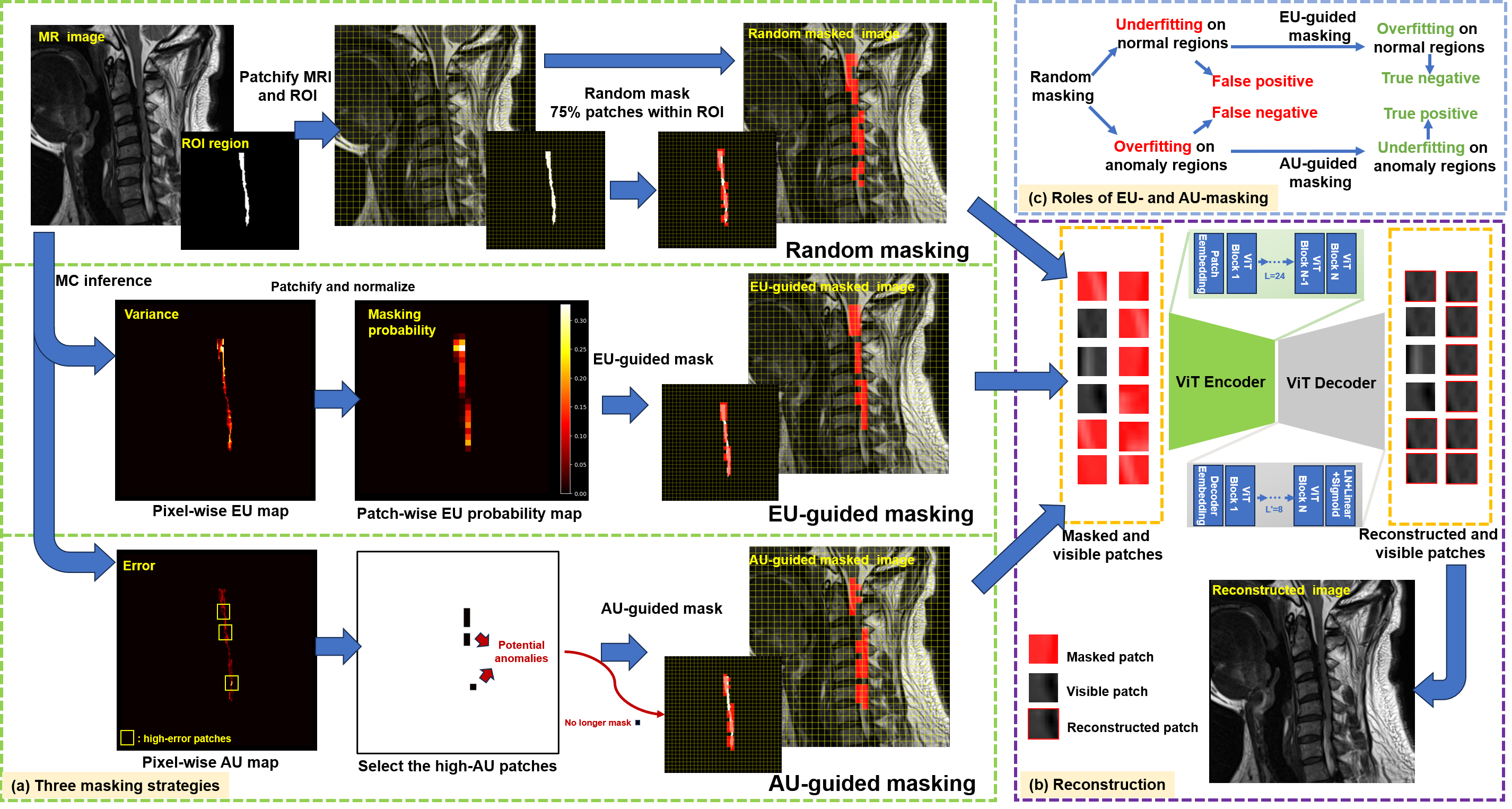}}
\caption{Comparison of three masking strategies in the $U^{2}AD$ framework: random masking, epistemic uncertainty (EU)-guided masking, and aleatoric uncertainty (AU)-guided masking. (a) The top-left panel illustrates the random masking strategy, where patches are randomly masked before being input into the ViT encoder-decoder for reconstruction. The middle panel demonstrates the proposed EU-guided masking strategy, leveraging variance maps from MC inference to generate masking probability for each patch, prioritizing low-confidence regions during training. The bottom panel presents the AU-guided masking strategy, where high-error patches in the error map (representing potential anomalies) are excluded from training, thus amplifying reconstruction errors in anomaly regions. (b) Masked patches from all three strategies are processed through the ViT encoder-decoder, resulting in reconstructed images that showcase differences in anomaly detection performance. (c) Highlights the roles of EU- and AU-guided masking in mitigating underfitting and overfitting, achieving balanced reconstruction for normal regions and effective anomaly detection.}
\label{strategy}
\end{figure*}

\subsection{Network Architecture}
\label{Network Architecture}
%The $U^{2}AD$ framework employs a ViT-based MAE with an asymmetric encoder-decoder structure. 
The encoder processes unmasked patches from the ROI, projecting them into an embedding space, while the decoder reconstructs the masked patches using embeddings from the encoder. Input images are divided into patches, and 75\% are masked during training. Positional encodings and a class token are added to the embeddings for processing through multi-head self-attention (MSA) and multi-layer perceptron (MLP) layers in the ViT. The decoder uses learnable tokens for masked patches, reconstructing pixel values and generating final outputs. A dual loss function, incorporating mean square error (MSE) loss and edge reconstruction loss, ensures the model focuses on both pixel accuracy and fine anatomical details, crucial for detecting small anomalies like T2 hyperintensities. In our study, both the pretraining and adaptation phases share the same network architecture.

\subsubsection{Encoder}
The input image \( x \in \mathbb{R}^{H \times W \times C} \), where \( H \), \( W \), and \( C \) denote the height, width, and number of channels, respectively, is divided into patches covering the ROI, specifically the spinal cord in this study. Let \( N \) represent the total number of patches within the ROI, and each patch \( x_p \in \mathbb{R}^{P^2 \times C} \), where \( P \) is the patch size. We randomly mask 75\% of the patches, leaving $ N' = 25\% \times N $ unmasked patches to be processed by the encoder.

Each unmasked patch \( x_{ip} \in \mathbb{R}^{P^2 \times C} \) is linearly projected into a 1D token \( x_{ip}^E \in \mathbb{R}^{1 \times D} \), where \( D \) is the embedding dimension. All unmasked tokens are concatenated to form the embedding space \( x_p^E \in \mathbb{R}^{N' \times D} \). To this embedding, we add a class token \( x_{\text{class}} \) and positional encoding \( E_{\text{pos}} \), resulting in the input embeddings:
\[
z_0 = [x_{\text{class}}; x_{1p}^E; x_{2p}^E; \dots; x_{N'p}^E] + E_{\text{pos}}.
\]

These embeddings are passed through \( L \) ViT blocks, where each block consists of a MSA mechanism and a MLP layer. The operations within each block \( l \) are as follows:
\[
z'_l = \text{MSA}(\text{LN}(z_{l-1})),
\]
where \( \text{LN}(\cdot) \) represents Layer Normalization, followed by:
\[
z_l = \text{MLP}(\text{LN}(z'_l)).
\]

The output \( z_L \) from the final encoder block represents the encoded embeddings of the unmasked patches, which are then passed to the decoder along with learnable tokens for the masked patches.

\subsubsection{Decoder}

The decoder reconstructs the masked patches using the embeddings from the unmasked patches output by the encoder. First, learnable tokens \( z_{\text{mask}} \in \mathbb{R}^{N_{\text{mask}} \times D} \), where \( N_{\text{mask}} = N - N' \), are introduced for the masked patches. These tokens are concatenated with the encoder output \( z_L \) and the class token \( x_{\text{class}} \), along with the positional encoding \( E_{\text{pos}} \), forming the decoder input:
\[
z_{\text{input}} = [x_{\text{class}}; z_L; z_{\text{mask}}] + E_{\text{pos}}.
\]

The input \( z_{\text{input}} \) is processed by \( L' \) ViT blocks, each with the same structure as the encoder blocks. For each block \( l' \), the operations are:
\[
z'_{l'} = \text{MSA}(\text{LN}(z_{l'-1})),
\]
followed by:
\[
z_{l'} = \text{MLP}(\text{LN}(z'_{l'})).
\]

The output \( z_{L'} \) from the final decoder block is passed through a linear projection layer \( f_{\text{proj}} \) to predict the pixel values of the masked patches. Let \( \hat{x}_p \in \mathbb{R}^{P^2 \times C} \) denote the reconstructed patch for each masked region. The final reconstruction is:
\[
\hat{x} = f_{\text{proj}}(z_{L'}).
\]

The reconstruction loss is computed by comparing the predicted masked patches \( \hat{x} \) with the ground truth non-masked images \( x\), enabling the model to learn the anatomy of the spinal cord.

\textbf{Loss Function:} We employ a dual reconstruction loss to train the model, inspired by Edge-MAE\citep{li2023multi}. The total loss function is defined as:
\[
L_{\text{Recon}} = L_{\text{MSE}} + \lambda L_{\text{Edge}},
\]

\[
L_{\text{MSE}} = \sum_{i \in \text{masked patches}} ||x_{ip} - \hat{x}_{ip}||_2,
\]

\[
L_{\text{Edge}} = \sum_{i \in \text{masked patches}} ||\text{Sobel}(x_{ip})-\text{Sobel}(\hat{x}_{ip})||_2.
\]
where \( L_{\text{Edge}} \) is the edge reconstruction loss, which is computed using Sobel edge detection. The edge reconstruction loss ensures that the model focuses on fine anatomical structures, such as spinal cord boundaries. The parameter \( \lambda \) controls the weight of the edge loss.

\subsection{Pretraining on Healthy Images}
\label{pretraining}
% 介绍原始的MAE用于自监督预训练，在本研究中的作用，以及存在的问题。一方面自监督MAE使用的是随机mask和重建作为pretext task，可以帮助ViT模型学习到更全面的特征，将图像有效地进行编码，另外在健康数据集上进行训练也有助于构建一个正常数据分布的隐向量空间，有助于在后续的异常检测中将异常作为OOD区分出来。但是MAE原始的mask-and-reconstruct的问题在于它是完全随机的，没有利用到更多的信息。
%Before $U^{2}AD$ is applied on target clinical datasets, it undergoes pretraining on a large, publicly available dataset of healthy spinal cord MR images using the random masking strategy, consistent with the MAE\cite{he2022masked}. The goal of pretraining is two-folds: (1) it acts as a pretext task that enables the ViT models to learn meaningful feature representations. Given that ViT models are known for being data-hungry, pretraining on a large dataset is crucial, particularly when the model is later applied to smaller clinical datasets; (2) training on healthy spinal cord images helps reconstruction model establish a latent manifold for normal anatomical distributions. This is essential for distinguishing anomalies as out-of-distribution samples in the further task.
%The pretrained model parameters \(\theta_{\text{pre}}\) are preserved for subsequent uncertainty estimation and adaptation training.

Before applying $U^{2}AD$ to clinical datasets, it is pretrained on a large dataset of healthy spinal cord MR images using random masking, as in MAE \citep{he2022masked}. Pretraining serves two purposes: (1) it enables ViT models to learn meaningful feature representations, essential given their data-hungry nature, and (2) it establishes a latent manifold for normal anatomical distributions, aiding in the identification of anomalies as out-of-distribution samples. The pretrained model parameters \(\theta_{\text{pre}}\) are preserved for subsequent uncertainty estimation and adaptation training.

However, random masking overlooks regional complexity, leading to two issues: high reconstruction errors in complex normal regions may cause false positives, while low reconstruction errors in anomalies due to generalization may result in false negatives. To address these, we propose EU- and AU-guided masking during adaptation training, prioritizing uncertain regions and enhancing differentiation between normal and abnormal areas. \Cref{strategy} illustrates the comparison between masking strategies.

\subsection{Uncertainty-Guided Masking and Adaptation Training}
\label{adaptation training}
%challenge可以不提，个人经验Method不能像discussion部分一样讨论前因后果然后再提出我们做了什么，博士论文这么写完全没问题，期刊论文在method部分直接摆出来做了什么，怎么做的就行，也给审稿人提问的空间
%In the adaptation training phase, the challenge lies in adapting the model to target data while avoiding overfitting on anomaly regions. The proposed Uncertainty-Guided Adaptation (UG-Ada) address these challenges by focusing on uncertain regions to refine reconstructions and enhance anomaly detection. 
To quantify both AU and EU maps, we propose a Monte-Carlo (MC)-based inference and uncertainty estimation method, leveraging the pretrained $U^{2}AD$ model. The detailed process is given in \Cref{uncertainty_estimation}. Regions with high EU indicate low model confidence and require prioritized adaptation, whereas regions with high AU reflect inherent abnormalities.

\normalem
\begin{algorithm}[H]
\caption{Monte Carlo Estimation of Aleatoric and Epistemic Uncertainty}\label{uncertainty_estimation}
\small
\SetAlgoLined
\KwIn{MAE model parameters $\theta$, test image \(x\), ROI mask \(M\), MC number \(K\)}
\KwOut{Aleatoric uncertainty map $\mathbf{AU}$, Epistemic uncertainty map $\mathbf{EU}$}

$\mathbf{x_p}=\{x_{1p}, x_{2p}, ..., x_{Np}\} \gets Patchify(x \odot M)$ \tcp{Extract ROI patches from image}
$\mathbf{C} \gets \{c_1 = 0, c_2 = 0, ..., c_N = 0\}$ \tcp{Initialize MC counters for each patch}
$\mathbf{R_p} \gets \{R_{1p}: [], R_{2p}: [], ..., R_{Np}: []\}$ \tcp{Initialize arrays to store reconstructions for each patch}

\While{$\min(\mathbf{C}) < K$}{

    $\mathbf{\overline{x}_p^{(I)}} = \{x_{1p}^{(I)}, x_{2p}^{(I)}, ..., x_{N'p}^{(I)}\}, N' = 75\% \times N \gets Random(\mathbf{x_p}, 75\%)$ \tcp{Randomly select 75\% patches for masking}
    
    $\mathbf{\tilde{x}_p^{(I)}} = \mathbf{x_p} \setminus \mathbf{\overline{x}_p^{(I)}}$ \tcp{Remaining 25\% visible patches}  
    $\mathbf{\hat{x}_p^{(I)}} = MAE_{\theta}(\mathbf{\tilde{x}_p^{(I)}})$ \tcp{Reconstruct masked patches using MAE}
    
    \For{$\hat{x}_{ip}^{(I)} \in \mathbf{\hat{x}_p^{(I)}}$}{
        $c_i \gets c_i + 1$ \tcp{Increment counter for selected patch}
        $R_{ip} \gets R_{ip} \cup \{\hat{x}_{ip}^{(I)}\}$ \tcp{Store the \(I\)-th reconstruction result for patch \(i\)}
    }
}

$\mathbf{AU} = \{AU_{1p}, AU_{2p}, ..., AU_{Np}\}$ \tcp{Initialize aleatoric uncertainty map}
$\mathbf{EU} = \{EU_{1p}, EU_{2p}, ..., EU_{Np}\}$ \tcp{Initialize epistemic uncertainty map}

\For{$x_{ip} \in \mathbf{x_p}$}{
    $R'_{ip} \gets RandomSelect(R_{ip}, K)$ \tcp{Randomly select \(K\) reconstructions for patch \(i\)}
    $\mu_{ip} \gets \frac{1}{K} \sum_{k=1}^{K} \hat{x}_{ip}^{(k)}$ \tcp{Compute mean reconstruction for patch \(i\)}
    $\sigma^2_{ip} \gets \frac{1}{K-1} \sum_{k=1}^{K} \left(\hat{x}_{ip}^{(k)} - \mu_{ip}\right)^2$ \tcp{Compute variance for patch \(i\)}
    
    $\mathbf{AU}[x_{ip}] \gets |x_{ip} - \mu_{ip}|$ \tcp{Aleatoric uncertainty: absolute error}
    $\mathbf{EU}[x_{ip}] \gets \sigma^2_{ip}$ \tcp{Epistemic uncertainty: variance}
}

\Return $\mathbf{AU}, \mathbf{EU}$
\end{algorithm}
\ULforem

Following pretraining, \textbf{Uncertainty-Guided Adaptation (UG-Ada)} are performed on target clinical datasets, which comprises two stages detailed in \Cref{UG-Ada}:
\begin{enumerate}
\item EU-Guided Adaptation: Highlights low-confidence regions for iterative reconstruction improvement.
\item AU-Guided Adaptation: Identifies and excludes high-error regions to amplify the distinction between normal and anomalous areas.
\end{enumerate}

\normalem
\begin{algorithm}[H]
\caption{Two-Stage Uncertainty-Guided Adaptation on Target Data}\label{UG-Ada}
\small
\SetAlgoLined
\KwIn{$U^{2}AD$ model parameters $\theta$, test image \(x\), ROI mask \(M\), epistemic map $\mathbf{EU}$, aleatoric map $\mathbf{AU}$, iteration interval $Q$, temperature $\tau$}
\KwOut{Updated model parameters $\theta_{\text{ada}}$}

$\mathbf{x_p} = \{x_{1p}, x_{2p}, ..., x_{Np}\} \gets Patchify(x \odot M)$ \tcp{Extract ROI patches from the masked image}
$\mathbf{P} = \{P_{1p}, P_{2p}, ..., P_{Np}\} \gets Random(0, 1)$ \tcp{Initialize random probabilities for each patch}

\While{Stage 1: EU-Guided Adaptation}{
    $\mathbf{EU}_{ip} = \sum_{(m,n) \in x_{ip}} \mathbf{EU}[x](m,n)$ \tcp{Compute patch-level EU values}
    $\mathbf{W}_{ip} = \frac{-\log(\exp(\mathbf{EU}_{ip} / \tau))}{\sum_{k=1}^{N} \exp(\mathbf{EU}_{kp} / \tau)}$ \tcp{Compute weighting factors for patches}
    $\mathbf{P'}_{ip} = \mathbf{W}_{ip} \cdot \mathbf{P}_{ip}$ \tcp{Weighted selection probabilities}

    $\mathbf{\overline{x}_p} = \{x_{ip} \mid \mathbf{P'}_{ip} \text{ in top 75\% of } \mathbf{P'}\}$ \tcp{Select top 75\% patches for masking}
    $\mathbf{\tilde{x}_p} = \mathbf{x_p} \setminus \mathbf{\overline{x}_p}$ \tcp{Visible patches (remaining 25\%)}
    $\mathbf{\hat{x}_p} \gets MAE_{\theta}(\mathbf{\tilde{x}_p})$ \tcp{Reconstruct masked patches using visible patches}
    Update $\theta$ using $\mathcal{L}_{reconstruction}$ \tcp{Update model parameters}

    \If{$epoch \mod Q == 0$}{
        $\mathbf{EU} \gets Update(\theta)$ \tcp{Refresh EU map every $Q$ epochs}
    }
}

\While{Stage 2: AU-Guided Adaptation}{
    $\text{AnoScore}_{CC} = \{\sum_{(m,n) \in CC} \mathbf{AU}[x](m,n) \mid CC \text{ is a connected component}\}$ \tcp{Calculate anomaly scores}
    $\text{TopCC} = \{CC_i \mid \text{AnoScore}_{CC_i} \text{ in top 3 of } \text{AnoScore}_{CC}\}$ \tcp{Select top 3 components}
    
    $\mathbf{P'}_{ip} = 0 \text{ for } x_{ip} \in \bigcup \text{TopCC}$ \tcp{Set probabilities to 0 for high-AU patches}
    $\mathbf{\overline{x}_p} = \{x_{ip} \mid \mathbf{P'}_{ip} \text{ in top 75\% of } \mathbf{P'}\}$ \tcp{Mask top 75\% patches excluding potential anomalies}
    $\mathbf{\tilde{x}_p} = \mathbf{x_p} \setminus \mathbf{\overline{x}_p}$ \tcp{Visible patches (remaining 25\%)}
    $\mathbf{\hat{x}_p} \gets MAE_{\theta}(\mathbf{\tilde{x}_p})$ \tcp{Reconstruct masked patches}
    Update $\theta$ using $\mathcal{L}_{reconstruction}$ \tcp{Update model parameters}

    \If{$epoch \mod Q == 0$}{
        $\mathbf{AU} \gets Update(\theta)$ \tcp{Refresh AU map every $Q$ epochs}
    }
}

$\theta_{\text{ada}} \gets \theta$ \tcp{Final adapted parameters}
\Return $\theta_{\text{ada}}$
\end{algorithm}
\ULforem

\subsubsection{Epistemic Uncertainty-Guided Adaptation}
In the first stage, EU-guided masking prioritizes regions where the model demonstrates low confidence. Instead of random masking, patches with higher EU are selected more frequently. The pixel-summation EU value for each patch within the ROI \(x_{ip}\) is computed using the corresponding EU map as follows:
\[
\text{EU}_{ip} = \sum_{(m,n) \in x_{ip}} \text{EU}[x](m,n),
\]

where \((m,n)\) indexes the pixels within the patch \(x_{ip}\). Patches with higher EU values are assigned a higher selection probability using a weighted adjustment:
\[
w_{ip} = \frac{-\log(\exp(\text{EU}_{ip}/\tau))}{\sum_{k=1}^{n} \exp(\text{EU}_{ip}/\tau)},
\]
where \(\tau\) is a temperature hyperparameter that controls the sharpness of the probability distribution. Adjusting \(\tau\) allows for tuning how strongly the model focuses on uncertain patches. The final weighted selection probability for each patch is calculated as:
\[
P'_{ip} = w_{ip} \times P'_{ip},
\]
Patches with the highest probabilities are masked for reconstruction, and uncertainty maps are updated every \(Q\) epochs. \(Q\) value is empirically selected as 10 to balance the performance and training time.

By focusing on high-EU regions, the model iteratively improves reconstruction performance in challenging areas, preventing underfitting and capturing complex anatomical details effectively

\subsubsection{Aleatoric Uncertainty-Guided Adaptation}
%In the EU-guided adaptation stage, anomalous regions are treated as normal, which can lead to the model generalizing to these abnormal representations. To address this limitation, 
AU-guided adaptation focuses on distinguishing and excluding anomalous areas from training. Patches with high AU values, indicative of reconstruction errors, are identified as potential anomalies. Anomaly scores for connected components (CCs) in the AU map are calculated as:
\[
\text{AnoScore}_{CC} = \sum_{(m,n) \in \text{CC}} \text{AU}[x](m,n),
\]

where \((m,n)\) indexes the pixels within it. The top three components with the highest scores are excluded from training to preserve their high reconstruction errors, ensuring anomalies remain distinguishable. Normal patches are trained as usual, improving reconstruction fidelity and further amplifying the contrast between normal and anomalous regions.

Uncertainty maps are updated iteratively during this stage, and the model parameters \(\theta_{\text{ada}}\) are refined. This dual-stage process effectively adapts the model to test data, generating refined anomaly maps that highlight abnormalities with high sensitivity.

\subsection{Postprocessing and Anomaly Detection}
\label{Postprocessing and Anomaly Detection}
\subsubsection{Anomaly Map Generation and Postprocessing}
The anomaly map is generated similarly to AU estimation. Using the adapted model parameters \(\theta_{\text{ada}}\), the ViT model performs Monte Carlo inference to obtain \(K\) reconstruction maps for each test image. The anomaly map for a test image \(x\) is computed as the absolute error between the original image and the average of the \(K\) reconstructions:
\[
\text{AnoMap}[x] = \left| x - \frac{1}{K} \sum_{k=1}^{K} \hat{x}^{(k)} \right|,
\]

In the Postprocessing step, \text{AnoMap}[x] is refined to focus on the most significant anomalies. First, a threshold filters out the bottom 20\% of error values, retaining prominent anomalous regions. Next, CCs within the refined anomaly map are identified and labeled. An anomaly score for each CC is computed as the summation of error values within the component. The top three CCs with the highest scores are retained, emphasizing the most prominent anomalies for further analysis.

\subsubsection{Error Signal Curve Plotting and Anomaly Detection}
To visualize anomalies along the spinal cord, an anomaly curve (AnoCurve) is computed by averaging anomaly values across the width of the spinal cord at each longitudinal position:

\[
AnoCurve(j) = \frac{1}{W} \sum_{i=1}^{W} \text{AnoMap}[x](i,j),
\]

where \(W\) is the spinal cord width and \(j\) is the longitudinal position. This curve provides a profile of anomalies along the spinal cord axis.

Anomalies are detected at both segment and patient levels:

\begin{itemize}
    \item \textbf{Segment-level Anomaly Score (\(AnoScore(S_i)\))}: Represents the peak anomaly value within each vertebral segment \(S_i\)(e.g., C2-C3):
    \begin{equation}
    AnoScore^S(S_i) = \max_{j \in S_i} AnoCurve(j).
    \end{equation}
    \item \textbf{Patient-level Anomaly Score (\(AnoScore(x)\))}: Represents the maximum anomaly value across the entire spinal cord:
    \begin{equation}
    AnoScore^P(x) = \max_{j} AnoCurve(j).
    \end{equation}
\end{itemize}

Thresholds on \(AnoScore^S(S_i)\) and \(AnoScore^P(x)\) are used to identify anomalous segments and patients. Segments exceeding the threshold are flagged for potential pathology, and high relative anomaly amplitudes indicate regions that may require further clinical investigation.

\section{Experiments and Results}

\begin{figure*}[!ht]
\centerline{\includegraphics[width=\columnwidth]{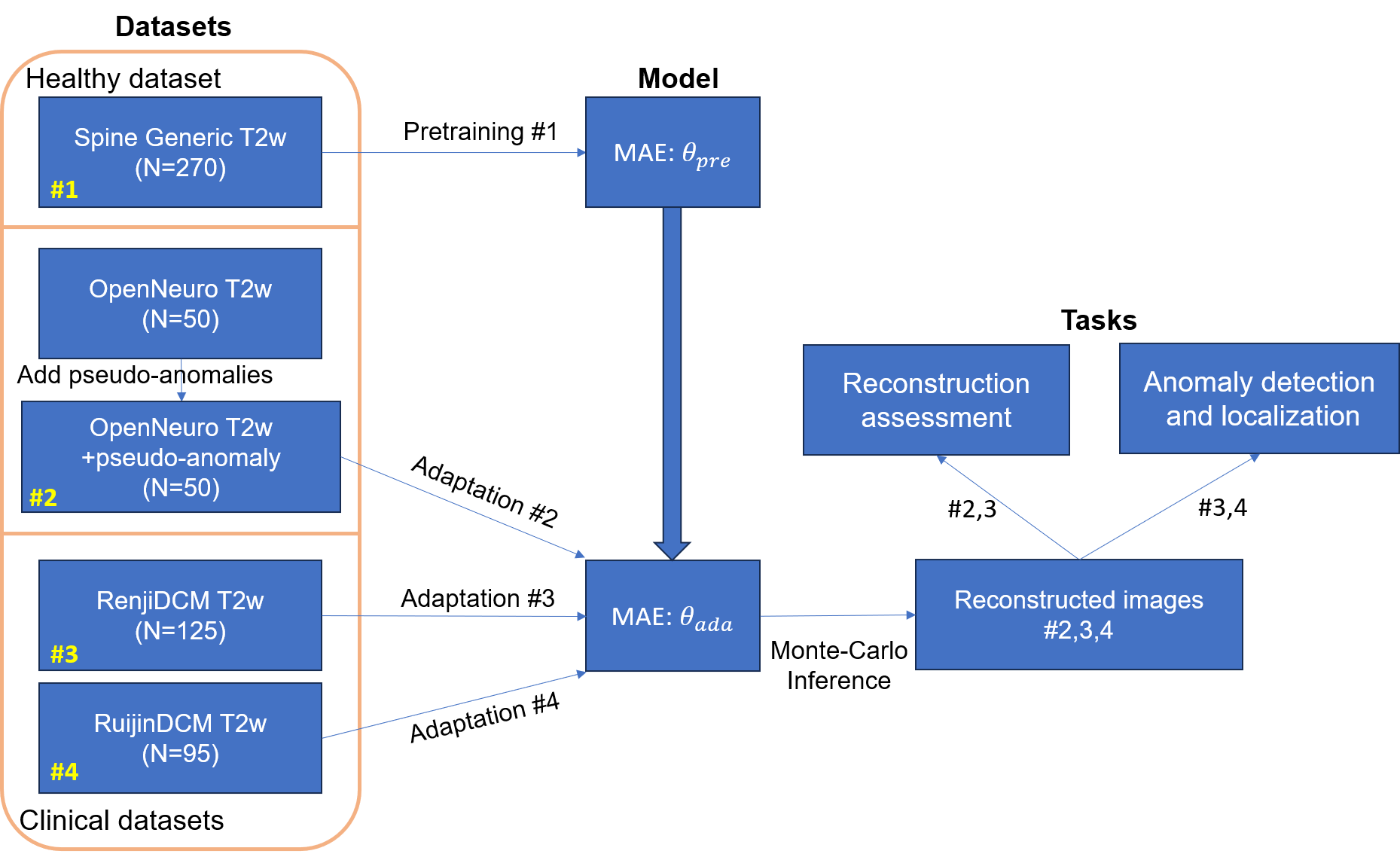}}
\caption{Datasets and tasks description for evaluating the $U^{2}AD$ framework. The Spine Generic T2w dataset (\#1) is used for pretraining the MAE model (\(\theta_{\text{pre}}\)). Adaptation training is applied on the OpenNeuro T2w dataset (\#2), RenjiDCM T2w dataset (\#3), and RuijinDCM T2w dataset (\#4). The OpenNeuro T2w dataset includes pseudo-anomalies and is designed for reconstruction assessment in anomaly-specific regions. The RenjiDCM T2w dataset is used to evaluate reconstruction performance across the entire ROI, while both RenjiDCM T2w (\#3) and RuijinDCM T2w (\#4) are evaluated for anomaly detection and localization tasks.}
\label{Datasets}
\end{figure*}

\subsection{Datasets and Tasks}
We collected datasets from multiple sites, including both publicly available and clinical datasets. \Cref{Datasets} shows the datasets and tasks in our study. The research aims to evaluate the models in two aspects: \textbf{Reconstruction Assessment} and \textbf{Anomaly Detection and Localization}. %For reconstruction task, we add an extra experiment on pseudo-anomalous dataset (OpenNeuro T2w \#2), which specifically evaluates the reconstruction results within anomalous regions. For anomaly detection task, we evaluate patient-level identification and segment-level localization on two clinical datasets.

%This task evaluates the reconstruction performance specifically on pseudo-anomalous regions. The reconstructed regions are compared with their corresponding original normal regions. \textbf{Task 2} is \textbf{Quantitative Evaluation on Whole ROI Regions}. This task assesses reconstruction performance over the entire ROI, encompassing both normal and abnormal regions. \textit{Task 3} is \textbf{Anomaly Detection}. This task focuses on detecting anomalies and includes two sub-tasks: patient-level identification and segment-level localization. Patient-level identification determines whether a patient has any anomalies, while segment-level localization identifies whether the detected anomalies are present in the specific segments. These two tasks are independent binary classification tasks and can be assessed using binary classification metrics.  
%The Spine Generic T2w (\#1) \cite{cohen2021open} is a public dataset, serving as the healthy public dataset in this study; OpenNeuro T2w (\#2) \cite{brennan2019longitudinal} is also a public dataset, but we have introduced pseudo-anomalies into it. We collected MR T2-weighted images from DCM patients at two hospitals, named RenjiDCM T2w (\#3) and RuijinDCM T2w (\#4), specifically for detecting T2 hyperintensities in the images. 
\begin{enumerate}
    \item \textbf{Spine Generic T2w (\#1) \citep{cohen2021open}}: This dataset is a large-scale multi-site public dataset containing 270 healthy spinal cord MR T2-weighted images. Each MR image has a resolution of 60×256×256, and the middle 30 sagittal slices from each 3D MR image are extracted to form the dataset, resulting in 8100 2D MR T2w images. This dataset serves the purpose of pretraining in this study and is not employed for anomaly detection tasks.
    \item \textbf{OpenNeuro T2w (\#2) \citep{brennan2019longitudinal}}: This dataset is a public dataset consisting of 50 3D MR T2-weighted images, each with a resolution of 64×250×250. We extract the middle 8 sagittal slices from each 3D MR image to create pseudo-anomalous images. The process of generating pseudo-anomalies is detailed in \ref{Generation_of_pseudo-anomalies}. The post-processed dataset consists of 400 cases, with each case containing three parts: the original image, the image with anomalies added, and the mask of the anomalous regions. This dataset is specifically designed to evaluate the reconstruction on anomalous regions.
    
    %The process of generating pseudo-anomalies involves creating both anomaly masks and anomalous signal distribution. The anomaly mask is elliptical in shape, with the width being a random value between 0.7 and 1 times the spinal cord width (w) at the placement location, and the length being a random value between 1 and 4 times the width. The pseudo-anomalous signal follows a Gaussian distribution. Initially, we compute the mean and variance of the signal distribution in the spinal cord and cerebrospinal fluid (CSF) regions. The T2 hyperintensity signal should have a higher intensity than the spinal cord but lower than the CSF, so the mean and variance of the generated pseudo-anomalous signal distribution are randomly selected between the mean values of the spinal cord and CSF. For each 2D image, we randomly generate 1 to 3 pseudo-anomalous regions and place them within the spinal cord region. The post-processed dataset consists of 400 cases, with each case containing three parts: the original image, the image with anomalies added, and the mask of the anomalous regions. This dataset is specifically designed to evaluate the reconstruction on anomalous regions.
    \item \textbf{RenjiDCM T2w (\#3)}: This dataset is a clinical dataset consisting of MR T2-weighted images from 125 DCM patients collected at Renji Hospital. Each MR image has a resolution of 11×256×256, and we use the middle 3 sagittal slices from each 3D MR image for training the reconstruction model. For the anomaly detection task, only the middle slices are used. All 2D images have been manually reviewed and annotated by doctors, with T2 hyperintensities labeled at the segment level(e.g., T2 hyperintensity between C4-5). Among the 125 images, 78 are labeled as containing T2 hyperintensities, while the remaining 47 are classified as normal. %It is worth noting that "normal" refers to images with normal spinal cord signal intensity, although other cervical spine pathologies may still exist and influence T2 hyperintensity detection.
    \item \textbf{RuijinDCM T2w (\#4)}: This clinical dataset consists of T2-weighted MR images from 95 DCM patients, collected at Ruijin Hospital. Each MR image has a resolution of 11×256×256. The data processing and task setup for this dataset mirror those of the RenjiDCM T2w dataset. However, this dataset is more challenging due to the extreme sparsity of anomalies: only 9 images are labeled as containing T2 hyperintensities, while the remaining 86 are classified as normal. The low prevalence of anomalies significantly increases the difficulty of detection.
\end{enumerate}

\subsection{Evaluation measures}

\textbf{Assessment of anomaly detection}: For patient-level detection, images are classified as anomalous or non-anomalous based on their maximum anomaly score, with evaluation metrics including \textbf{accuracy, F1 score, and specificity}. To prioritize clinical sensitivity, the anomaly score threshold is set to achieve a sensitivity of 0.9, ensuring most T2 hyperintensities are detected. For segment-level localization, the segment with the highest anomaly score is classified as anomalous or non-anomalous, evaluated using \textbf{F1 score, recall, and specificity}. The segment-level anomaly score threshold is optimized to maximize the F1 score on the test set.
%For patient-level identification, images are classified as anomalous or non-anomalous based on the maximum anomaly score in the image. \textbf{Accuracy, F1 score, and specificity} are reported for evaluating this binary classification task. Considering the critical need for high sensitivity in clinical applications, the anomaly score threshold is set such that the sensitivity of the test set reaches 0.9. This ensures that most T2 hyperintensities are detected. For segment-level localization, the segment with the highest anomaly score is selected and classified as either anomalous or non-anomalous. \textbf{F1 score, recall, and specificity} are used to evaluate this task. The anomaly score threshold for segment-level is determined by selecting the value that maximizes the F1-score on the test set.

\textbf{Assessment of reconstruction}: Reconstruction performance is evaluated using standard metrics, including \textbf{Peak Signal-to-Noise Ratio (PSNR)}, \textbf{Structural Similarity Index Measure (SSIM)}, and \textbf{Mean Squared Error (MSE)}. To assess model confidence in the proposed $U^{2}AD$ framework, \textbf{variance} is introduced as an additional metric in the ablation study.
%The objective of the reconstruction task contrasts with that of anomaly detection. The reconstruction task aims to minimize reconstruction error, whereas the anomaly detection task seeks to maximize it. Specifically, we seek to minimize reconstruction error in normal regions and increase it in anomalous regions, which helps better distinguish between normal and abnormal areas. 
%For evaluating reconstruction in normal regions, we use standard reconstruction metrics such as \textbf{PSNR, SSIM, and MSE} to calculate the results in the target image ROI. For anomalous region reconstruction, we aim to bring the reconstruction as close as possible to its normal state. Since obtaining paired normal images for the anomalous ones is challenging, we use the generated pseudo-anomalous image dataset and corresponding anomaly masks to assess the reconstruction of anomalous regions. We treat the initial image as the ground truth, perform reconstruction on the image with pseudo-anomalies added, and compute reconstruction metrics such as \textbf{PSNR, SSIM, and MSE} between the reconstructed pseudo-anomalous regions and the ground truth. 
%Reconstruction performance is evaluated using standard metrics including \textbf{PSNR, SSIM, and MSE}. To evaluate model confidence in the proposed $U^{2}AD$ framework, \textbf{variance} is introduced as an additional metric in the ablation study.

\subsection{Implementation Details}

\textbf{Image Pre-processing}: In this experiment, we begin by pre-processing all images in the dataset and segmenting the ROI regions. All experiments use 2D images, where sagittal slices of the MR volume are treated as separate 2D images. All 2D images are first resampled to a voxel size of 1mm × 1mm and then centrally cropped to a size of 256 × 256 to match the model’s input size. The ROI for this task is the spinal cord in each image slice, and we use a pre-trained 2D nnU-Net\citep{isensee2021nnu} to segment the spinal cord in each 2D image. Furthermore, to enable anomaly localization, we segment and label each cervical vertebra to associate detected anomalies with specific segments, ranging from C2-3 to C7-T1. 
%这里可以省略，审稿人问了再解释
%It should be noted that in the edge slices of each MR image, the spinal cord may fade or become indistinct, particularly in patients with cervical scoliosis. As a result, for the clinical datasets, we only annotate anomalies in the middle sagittal slice of each MR image, and anomaly detection is carried out only on these 2D slices. 
All images are normalized using min-max normalization to scale the signal intensity to a range of 0-1. 

\textbf{Model Training and Evaluation}: Our pre-training, adaptation, and inference all use the same architecture, EdgeMAE\citep{li2023multi}. EdgeMAE is built on the original MAE structure, with two output heads at the final decoder layer: one for the reconstructed image and the other for the edge map of the reconstruction. The encoder and decoder adopt 24 and 8 ViT blocks, respectively. The model undergoes 200 epochs of both pre-training and adaptive training. We set an initial learning rate of $3e^{-3}$, decreasing it by a factor of 0.1 every 50 epochs. The Adam optimizer was used with a batch size of 8, and $\beta_1$ set to 0.9 and $\beta_2$ to 0.95. We set the patch size to 8 based on empirical values, as it better matches the width of the spinal cord. A larger patch size would mask regions outside the spinal cord, while a smaller patch size would increase training and inference time, and might fail to cover the entire anomalous region. During training, data augmentation is applied by adding Gaussian noise with a mean of 0 and a variance of 0.02 to the images. The brightness and contrast of the images are also randomly adjusted between 0.8 and 1.2 times the original values. The images are then re-normalized to the range of 0-1.
%In the uncertainty generation process, our Monte Carlo inference aims to produce \(K\) reconstruction results. However, the actual inference process runs slightly more than \(K\) times because only a subset of patches is randomly selected in each inference, ensuring that every patch is selected and reconstructed at least \(K\) times.
We set the loss function weight $\lambda$ to 0.1. All experiments were conducted on an NVIDIA 4090 GPU with 24GB of memory. For the anomaly detection task, we performed 5-fold cross-validation. We divided the data into five folds, using four folds to find the optimal anomaly score threshold and the remaining fold to compute evaluation metrics. 

\subsection{Comparison Methods}

\textbf{UAD methods:} we compared our proposed method with several state-of-the-art UAD approaches, including VAE \citep{kingma2013auto}, DAE \citep{kascenas2022denoising}, VAEGAN \citep{larsen2016autoencoding}, f-AnoGAN \citep{schlegl2019f}, and AnoDDPM \citep{Wyatt_2022_CVPR}. We implemented these methods according to the original papers and codes, while adjusting the data input and training settings to suit our task, as illustrated in \ref{Details_of_UAD_methods}. Pretraining and adaptation training are set as 200 epochs for each, consistent with our method, ensuring that the models converged. For all UAD models, the post-processing and evaluation methods were consistent with those used in our proposed approach.

\textbf{Other methods:} To thoroughly evaluate the superiority of our method in T2 hyperintensity detection, we also compared it with supervised object detection methods and signal analysis methods. For object detection-based methods, we chose YOLOv8 \citep{varghese2024yolov8}, Mask R-CNN \citep{he2017mask}, and DETR \citep{zhu2020deformable}, all of which are end-to-end models. For signal analysis methods, we selected Z-scoring, Simple Linear Iterative Clustering (SLIC) and Density-Based Spatial Clustering of Applications with Noise (DBSCAN). Z-scoring is a statistical approach which is based on the standard normal distribution. SLIC and DBSCAN are clustering approaches. The setting details are listed in \ref{Details_of_Object_detection_methods} and \ref{Details_of_Signal_analysis_methods}.

\begin{table*}[!ht]
\centering
\caption{Comparison of UAD methods on anomaly detection tasks, including patient-level identification and segment-level localization. Models are evaluated on RenjiDCM T2w dataset and RuijinDCM T2w dataset.}
\label{Comparison of UAD Methods}
\scalebox{0.6}{
\begin{tabular}{|c|c|c|c|c|c|c|c|}
\hline
\multicolumn{8}{|c|}{\textbf{on RenjiDCM T2w dataset}}\\ \hline
\multicolumn{2}{|c|}{\multirow{2}{*}{\textbf{Models}}}& \multicolumn{3}{c|}{Patient-level identification} & \multicolumn{3}{c|}{Segment-level localization} \\ \cline{3-8} 
                       \multicolumn{2}{|c|}{} & Accuracy ↑          & F1 score ↑         & Specificity ↑      & F1 score ↑         & Recall ↑           & Specificity ↑      \\ \hline
\multirow{6}{*}{\parbox{1.7cm}{Strategy 1}} 
                       & VAE                 & 0.6320 ± 0.0531    & 0.7521 ± 0.0498    & 0.1667 ± 0.1363    & 0.4951 ± 0.1019    & 0.8600 ± 0.3130    & 0.1200 ± 0.2683    \\ 
                       & DAE                 & \uline{0.6480 ± 0.0588}    & \uline{0.7642 ± 0.0307}    & 0.2133 ± 0.1759    & \uline{0.5893 ± 0.0394}    & \textbf{0.9800 ± 0.0447}    & 0.0410 ± 0.0602    \\ 
                       & VAEGAN              & 0.6000 ± 0.0620    & 0.7344 ± 0.0549    & 0.1067 ± 0.0705    & 0.4452 ± 0.0225    & 0.9714 ± 0.0639    & 0.0333 ± 0.0745    \\ 
                       & f-AnoGAN            & 0.6400 ± 0.0253    & 0.7596 ± 0.0095    & 0.1911 ± 0.1076    & 0.3813 ± 0.0223    & 0.8143 ± 0.1308    & 0.1614 ± 0.1399    \\ 
                       & AnoDDPM             & \uline{0.6480 ± 0.0392}    & 0.7610 ± 0.0250    & 0.2356 ± 0.1112    & 0.5536 ± 0.1451    & 0.6071 ± 0.2406    & \textbf{0.7699 ± 0.0422}    \\ 
                       & $U^{2}AD$ (Ours)         & 0.6260 ± 0.0197    & 0.7411 ± 0.0463    & \uline{0.2574 ± 0.1596}    & 0.5770 ± 0.1084    & 0.6351 ± 0.2028    & 0.6244 ± 0.2585    \\ \hline
\multirow{6}{*}{\parbox{1.7cm}{Strategy 2}} 
                       & VAE                 & 0.6320 ± 0.1055    & 0.7419 ± 0.0992    & 0.2111 ± 0.0605    & 0.3400 ± 0.1786    & 0.4467 ± 0.2805    & \uline{0.6942 ± 0.1106}    \\ 
                       & DAE                 & 0.5920 ± 0.0299    & 0.7316 ± 0.0308    & 0.0822 ± 0.0756    & 0.4016 ± 0.0736    & 0.7607 ± 0.2777    & 0.1935 ± 0.2426    \\ 
                       & VAEGAN              & 0.6400 ± 0.0438    & 0.7617 ± 0.0307    & 0.1711 ± 0.0888    & 0.3187 ± 0.0215    & 0.9200 ± 0.1095    & 0.0886 ± 0.1075    \\ 
                       & f-AnoGAN            & 0.6240 ± 0.0742    & 0.7478 ± 0.0544    & 0.1733 ± 0.0923    & 0.4305 ± 0.0457    & \uline{0.9714 ± 0.0639}    & 0.0883 ± 0.0846    \\
                       & AnoDDPM             & 0.6240 ± 0.0697    & 0.7466 ± 0.0538    & 0.1711 ± 0.1046    & 0.2798 ± 0.1915    & 0.5300 ± 0.3194    & 0.5614 ± 0.1902    \\
                       & $U^{2}AD$ (Ours)         & \uline{0.7416 ± 0.0186}    & \uline{0.8126 ± 0.0520}    & \uline{0.4784 ± 0.1194}    & \uline{0.7068 ± 0.0741}    & 0.7892 ± 0.1112    & 0.6223 ± 0.1556    \\ \hline
\multirow{6}{*}{\parbox{1.7cm}{Strategy 3}} 
                       & VAE                 & 0.6880 ± 0.0688    & 0.7819 ± 0.0589    & 0.3200 ± 0.0959    & 0.5982 ± 0.0845    & 0.6429 ± 0.0970    & \uline{0.7667 ± 0.0850}    \\ 
                       & DAE                 & 0.6400 ± 0.0358    & 0.7590 ± 0.0272    & 0.1911 ± 0.0412    & 0.4049 ± 0.0495    & 0.7048 ± 0.1454    & 0.4140 ± 0.1699    \\ 
                       & VAEGAN              & 0.5920 ± 0.0588    & 0.7309 ± 0.0516    & 0.0867 ± 0.0827    & 0.2965 ± 0.1105    & 0.4533 ± 0.3532    & 0.6068 ± 0.3075    \\
                       & f-AnoGAN            & 0.6480 ± 0.0299    & 0.7555 ± 0.0386    & 0.2511 ± 0.1188    & 0.5540 ± 0.0656    & 0.8618 ± 0.1523    & 0.1210 ± 0.1090    \\ 
                       & AnoDDPM             & 0.6320 ± 0.0466    & 0.7519 ± 0.0350    & 0.1911 ± 0.0755    & 0.4796 ± 0.0992    & 0.4796 ± 0.0992    & 0.2912 ± 0.1801    \\
                       & $U^{2}AD$ (Ours)         & \textbf{0.7740 ± 0.0124}    & \textbf{0.8336 ± 0.0348}    & \textbf{0.5533 ± 0.1384}    & \textbf{0.7345 ± 0.0547}    & \uline{0.8829 ± 0.0977}    & 0.5298 ± 0.1379    \\ \hline
\multicolumn{8}{|c|}{\textbf{on RuijinDCM T2w dataset}}\\ \hline
\multicolumn{2}{|c|}{\multirow{2}{*}{\textbf{Models}}}& \multicolumn{3}{c|}{Patient-level identification} & \multicolumn{3}{c|}{Segment-level localization} \\ \cline{3-8} 
                       \multicolumn{2}{|c|}{} & Accuracy ↑          & F1 score ↑         & Specificity ↑      & F1 score ↑         & Recall ↑           & Specificity ↑      \\ \hline
\multirow{6}{*}{\parbox{1.7cm}{Strategy 1}} & VAE        & 0.1895 ± 0.0976 & 0.1698 ± 0.0397 & 0.1170 ± 0.1237 & 0.1500 ± 0.2236 & \uline{0.4000 ± 0.5477} & 0.8092 ± 0.0925 \\
                                           & DAE        & \uline{0.3579 ± 0.2455} & 0.2356 ± 0.0928 & \uline{0.3046 ± 0.2900} & 0.1816 ± 0.1776 & \uline{0.4000 ± 0.4183} & 0.8052 ± 0.1195 \\
                                           & VAEGAN     & 0.1895 ± 0.0421 & 0.1709 ± 0.0487 & 0.1170 ± 0.0382 & 0.2000 ± 0.2981 & 0.2000 ± 0.2739 & 0.9314 ± 0.0940 \\
                                           & f-AnoGAN   & 0.3263 ± 0.1644 & 0.2029 ± 0.0468 & 0.2693 ± 0.1965 & 0.0733 ± 0.1011 & \uline{0.4000 ± 0.5477} & 0.4889 ± 0.1040 \\
                                           & AnoDDPM    & 0.2632 ± 0.0744 & \uline{0.2632 ± 0.0744} & 0.1987 ± 0.0618 & \uline{0.3333 ± 0.4714} & 0.3333 ± 0.4714 & 0.3333 ± 0.4714 \\
                                           & $U^{2}AD$ (Ours) & 0.1474 ± 0.0624 & 0.1423 ± 0.0693 & 0.0802 ± 0.1313 & 0.2796 ± 0.3143 & 0.3350 ± 0.3866 & \uline{0.9753 ± 0.0374} \\
\hline
\multirow{6}{*}{\parbox{1.7cm}{Strategy 2}} & VAE        & 0.1684 ± 0.1021 & 0.1479 ± 0.0365 & 0.1059 ± 0.1310 & 0.0333 ± 0.0745 & 0.2000 ± 0.4472 & 0.8778 ± 0.2434 \\
                                           & DAE        & 0.6316 ± 0.0744 & 0.3129 ± 0.0755 & 0.6059 ± 0.0941 & 0.1600 ± 0.3578 & 0.2000 ± 0.4472 & 0.9092 ± 0.0844 \\
                                           & VAEGAN     & 0.2316 ± 0.1780 & 0.1805 ± 0.0355 & 0.1634 ± 0.2185 & 0.0667 ± 0.1491 & 0.2000 ± 0.4472 & 0.9222 ± 0.1083 \\
                                           & f-AnoGAN   & 0.4737 ± 0.0744 & 0.2406 ± 0.0607 & 0.4314 ± 0.0920 & 0.0667 ± 0.1491 & 0.2000 ± 0.4472 & 0.9333 ± 0.0913 \\
                                           & AnoDDPM    & 0.2000 ± 0.0516 & 0.2000 ± 0.0516 & 0.1275 ± 0.0421 & 0.1000 ± 0.2236 & 0.2000 ± 0.4472 & 0.8918 ± 0.0981 \\
                                           & $U^{2}AD$ (Ours) & \uline{0.6826 ± 0.1934} & \uline{0.3740 ± 0.2104} & \uline{0.6694 ± 0.2950} & \uline{0.5416 ± 0.3800} & \uline{0.5700 ± 0.4074} & \uline{0.9831 ± 0.0308} \\
\hline
\multirow{6}{*}{\parbox{1.7cm}{Strategy 3}} & VAE        & 0.4211 ± 0.1200 & 0.1797 ± 0.1006 & 0.3843 ± 0.1739 & 0.4333 ± 0.4346 & 0.5000 ± 0.5000 & 0.9654 ± 0.0317 \\
                                           & DAE        & 0.3684 ± 0.0333 & 0.2074 ± 0.0525 & 0.3144 ± 0.0491 & 0.0571 ± 0.1278 & 0.1000 ± 0.2236 & 0.9307 ± 0.0969 \\
                                           & VAEGAN     & 0.2526 ± 0.0698 & 0.1838 ± 0.0553 & 0.1869 ± 0.0700 & 0.0400 ± 0.0894 & 0.2000 ± 0.4472 & 0.8889 ± 0.1884 \\
                                           & f-AnoGAN   & 0.4211 ± 0.1912 & 0.2382 ± 0.0650 & 0.3732 ± 0.2324 & 0.2971 ± 0.3317 & 0.6000 ± 0.5477 & 0.8203 ± 0.0973 \\
                                           & AnoDDPM    & 0.4211 ± 0.1372 & 0.2280 ± 0.0592 & 0.3739 ± 0.1628 & 0.0800 ± 0.1789 & 0.2000 ± 0.4472 & 0.9556 ± 0.0724 \\
                                           & $U^{2}AD$ (Ours) & \textbf{0.7740 ± 0.0124} & \textbf{0.5569 ± 0.2333} & \textbf{0.8556 ± 0.1577} & \textbf{0.6846 ± 0.3084} & \textbf{0.7000 ± 0.3666} & \textbf{0.9857 ± 0.0270} \\
\hline
\multicolumn{8}{l}{Strategy 1: only train on public healthy data} \\
\multicolumn{8}{l}{Strategy 2: only train on clinical data}\\
\multicolumn{8}{l}{Strategy 3: pretrain on public healthy data and adaptation on clinical data}\\
\multicolumn{8}{l}{\textbf{Bold} values indicate the best performance for each metric, and \underline{underlined} values indicate the best for the strategy.} \\
\multicolumn{8}{l}{↑ indicates the higher value is better.} 
\end{tabular}
}
\end{table*}

\begin{table*}[!ht]
\centering
\caption{Comparison results on anomaly detection tasks of other methods, such as object detection-based and signal analysis-based models. Models are evaluated on RenjiDCM T2w dataset and RuijinDCM T2w dataset.}
\label{Comparison of different methods}
\scalebox{0.6}{
\begin{tabular}{|c|c|c|c|c|c|c|c|}
\hline
\multicolumn{8}{|c|}{\textbf{on RenjiDCM T2w dataset}}\\ \hline
\multicolumn{2}{|c|}{\multirow{2}{*}{\textbf{Models}}} & \multicolumn{3}{c|}{Patient-level identification} & \multicolumn{3}{c|}{Segment-level localization} \\ \cline{3-8} 
                        \multicolumn{2}{|c|}{} & Accuracy ↑         & F1 score ↑         & Specificity ↑       & F1 score ↑         & Recall ↑           & Specificity ↑      \\ \hline
\multirow{3}{*}{\parbox{1.5cm}{Object\\detection-based}} 
                        & YOLOv8              & 0.6764 ± 0.0579    & 0.7545 ± 0.0423    & 0.4689 ± 0.1804     & 0.7103 ± 0.0282    & 0.7093 ± 0.1073    & 0.5253 ± 0.2537    \\
                        & Mask R-CNN          & 0.6314 ± 0.0563    & 0.7540 ± 0.0389    & 0.1533 ± 0.2181     & 0.6984 ± 0.0399    & 0.8038 ± 0.0619    & 0.1765 ± 0.1121    \\ 
                        & DETR                & 0.6772 ± 0.0532    & 0.7800 ± 0.0283    & 0.2620 ± 0.2334     & 0.7332 ± 0.0818    & 0.8020 ± 0.1610    & 0.3595 ± 0.2100    \\ \hline
\multirow{3}{*}{\parbox{1.5cm}{Signal\\analysis-based}} 
                        & Z-scoring           & 0.6320 ± 0.0392    & 0.7522 ± 0.0291    & 0.1911 ± 0.0755     & 0.6621 ± 0.0684    & 0.8533 ± 0.1164    & 0.5642 ± 0.1132    \\ 
                        & SLIC                & 0.6480 ± 0.0299    & 0.7253 ± 0.0148    & 0.2328 ± 0.1234     & 0.5736 ± 0.1092    & 0.5778 ± 0.1826    & \textbf{0.7750 ± 0.0948}    \\ 
                        & DBSCAN              & 0.6560 ± 0.0599    & 0.6987 ± 0.0210    & 0.2196 ± 0.0591     & 0.5868 ± 0.0515    & 0.6956 ± 0.0482    & 0.5942 ± 0.1652    \\ \hline
\multicolumn{2}{|c|}{$U^{2}AD$(Ours)}  & \textbf{0.7740 ± 0.0124}    & \textbf{0.8336 ± 0.0348}    & \textbf{0.5533 ± 0.1384}     & \textbf{0.7345 ± 0.0547}    & \textbf{0.8829 ± 0.0977}    & 0.5299 ± 0.1379    \\ \hline
\multicolumn{8}{|c|}{\textbf{on RuijinDCM T2w dataset}}\\ \hline
\multicolumn{2}{|c|}{\multirow{2}{*}{\textbf{Models}}} & \multicolumn{3}{c|}{Patient-level identification} & \multicolumn{3}{c|}{Segment-level localization} \\ \cline{3-8} 
                        \multicolumn{2}{|c|}{} & Accuracy ↑         & F1 score ↑         & Specificity ↑       & F1 score ↑         & Recall ↑           & Specificity ↑      \\ \hline
\multirow{3}{*}{\parbox{1.5cm}{Object\\detection-based}} 
                        & YOLOv8              & 0.7300 ± 0.1271    & 0.2100 ± 0.2745    & 0.7300 ± 0.1301    & 0.2600 ± 0.2408    & 0.4000 ± 0.4183    & 0.8893 ± 0.064 \\
                        & Mask R-CNN          & 0.0238 ± 0.0178    & 0.0461 ± 0.034     & 0.0000 ± 0.0000    & 0.1714 ± 0.2556    & 0.4000 ± 0.5477    & 0.7949 ± 0.1071 \\
                        & DETR                & 0.0849 ± 0.0547    & 0.1528 ± 0.092     & 0.0000 ± 0.0000    & 0.2133 ± 0.307     & 0.2000 ± 0.2739    & 0.7383 ± 0.1769 \\
\hline
\multirow{3}{*}{\parbox{1.5cm}{Signal\\analysis-based}} 
                        & Z-scoring           & 0.4526 ± 0.1356    & 0.2380 ± 0.0560    & 0.4085 ± 0.1646    & 0.2356 ± 0.2186    & 0.5000 ± 0.5000    & 0.8281 ± 0.0911 \\
                        & SLIC                & 0.4526 ± 0.0714    & 0.2365 ± 0.0752    & 0.4078 ± 0.0684    & 0.3133 ± 0.3015    & 0.6000 ± 0.5477    & 0.9222 ± 0.0633 \\
                        & DBSCAN              & 0.5263 ± 0.0942    & 0.2605 ± 0.0546    & 0.4889 ± 0.1226    & 0.2000 ± 0.2981    & 0.3000 ± 0.4472    & 0.9425 ± 0.0721 \\
\hline
\multicolumn{2}{|c|}{$U^{2}AD$(Ours)}  & \textbf{0.7740 ± 0.0124}    & \textbf{0.5569 ± 0.2333}    & \textbf{0.8556 ± 0.1577}    & \textbf{0.6846 ± 0.3384}    & \textbf{0.7000 ± 0.3666}    & \textbf{0.9857 ± 0.0270} \\
\hline
\multicolumn{8}{l}{\textbf{Bold} values indicate the best performance for each metric.} \\
\multicolumn{8}{l}{↑ indicates the higher value is better.} 
\end{tabular}
}
\end{table*}

\begin{figure*}[!t]
\centerline{\includegraphics[width=0.7\columnwidth]{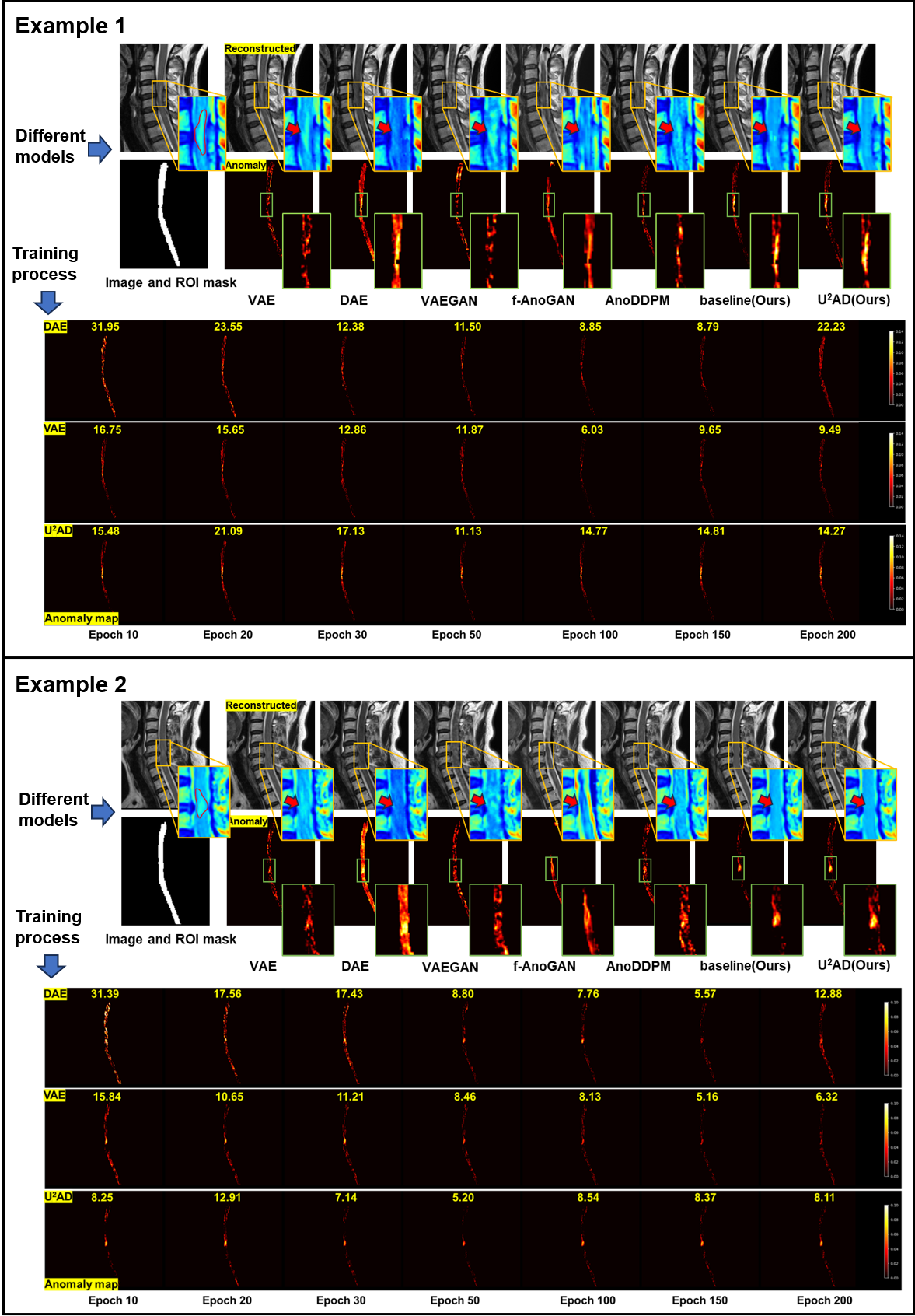}}
\caption{Examples of reconstruction results and anomaly maps generated by different UAD models. For each image, the reconstruction is limited within the ROI mask (the spinal cord). For each example, we compare the reconstruction results and anomaly maps between these UAD models. The true anomaly regions are highlighted with yellow boxes and red arrows, and magnified to better illustrate the results. Also, we present the varied anomaly maps during the training process for DAE, VAE, and $U^{2}AD$. The numerical values above each anomaly map represent the reconstruction error for the entire ROI.}
\label{UAD_anomaly_comparison}
\end{figure*}

\begin{figure*}[!t]
\centerline{\includegraphics[width=0.7\columnwidth]{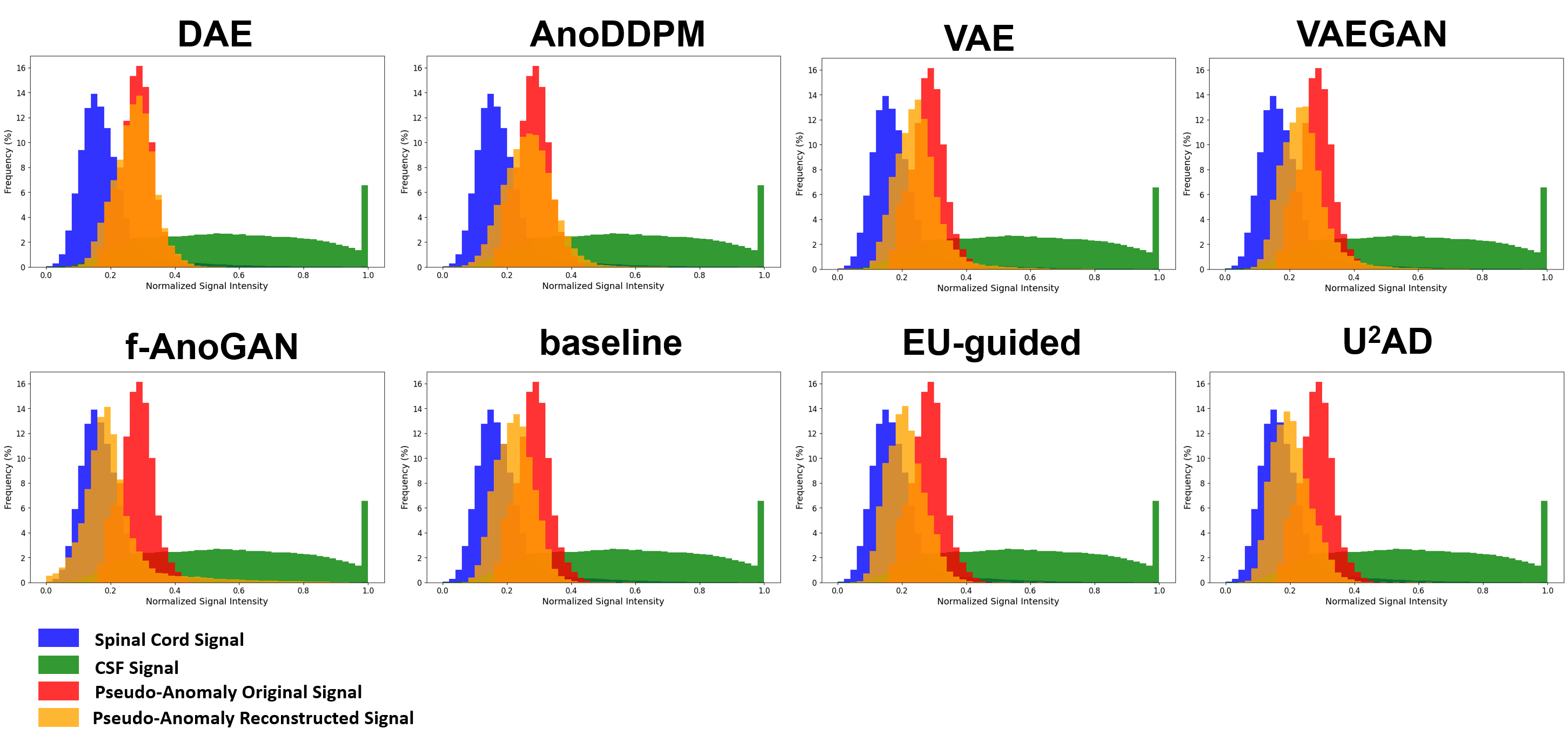}}
\caption{Signal intensity distributions of different regions in the pseudo-anomaly OpenNeuro T2w dataset. Blue and green represent the normal signal distributions of SC and CSF, respectively, while red indicates the pseudo-anomaly regions. The orange represents the reconstructed signal distribution of pseudo-anomaly regions from different UAD models.}
\label{openneuro_signaldist}
\end{figure*}

\subsection{Comparison Experiments with UAD Methods}
In this study, we not only evaluate the anomaly detection performance of different UAD methods, but also compare the effects of different dataset training strategies on each UAD method. We considered three dataset training strategies: Strategy 1: only pretrain on public healthy data without adaptation;  Strategy 2: only train on clinical data without pretraining; and  Strategy 3: pretraining and adaptation. It is important to note that the public healthy data and clinical data are from different distributions in this study, which introduces domain shift. However, the anomaly detection is only performed on the clinical data. The purpose of comparing these three training strategies is twofold: first, to investigate the impact of domain shift on unsupervised anomaly detection performance, and second, to identify the best dataset training strategy.

\Cref{Comparison of UAD Methods} presents the patient-level and segment-level anomaly detection results across different UAD models and dataset training strategies on the RenjiDCM T2w and RuijinDCM T2w datasets. Our proposed method, $U^{2}AD$, demonstrated the best overall performance under strategy 3. On the RenjiDCM T2w dataset, $U^{2}AD$ achieved the highest accuracy (0.7740), F1 score (0.8336), and specificity (0.5533) in the patient-level identification task. Notably, the F1 score and specificity represent a significant improvement compared to other models, where specificity values were mostly below 0.5, with several models scoring between 0.2 and 0.3. This highlights $U^{2}AD$'s ability to maintain high sensitivity while minimizing false positives. In the segment-level localization task, $U^{2}AD$ achieved the highest F1 score of 0.7345, outperforming all other models. While DAE under strategy 1 achieved a high recall of 0.9800, its specificity was very low (0.0410), indicating a tendency toward false positives. Conversely, AnoDDPM with strategy 1 achieved the highest specificity (0.7699) but exhibited poor recall (0.6071). $U^{2}AD$ achieved an optimal balance between specificity and recall, leading to its superior F1 score.

On the RuijinDCM T2w dataset, which is a screening-oriented dataset characterized by a low proportion of anomalies and predominantly normal cases, most UAD models, except for our proposed $U^{2}AD$, performed poorly in both patient-level identification and segment-level localization, with results indicating near failure. In contrast, $U^{2}AD$ demonstrated remarkable robustness under strategy 3, achieving the best performance across all metrics. For patient-level identification, $U^{2}AD$ achieved an F1 score of 0.5569, specificity of 0.8556, and accuracy of 0.7740, effectively distinguishing the small number of anomalous cases from the overwhelming majority of normal cases. In the segment-level localization task, $U^{2}AD$ further outperformed other models, attaining an F1 score of 0.6846 and a specificity of 0.9857, while maintaining high sensitivity.

%\Cref{performance_comparison} further visualizes the impact of three dataset training strategies on the performance of the six UAD models. The best dataset training strategy for each UAD model varies. From the results of RenjiDCM T2w dataset, VAE and $U^{2}AD$ performed best with strategy 3, while DAE, f-AnoGAN, and AnoDDPM performed best with strategy 1. VAEGAN performed best using strategy 2. We will discuss the reasons for these findings in the Discussion section. It is worth noting that $U^{2}AD$ did not achieve the best results in all three training strategies. When using strategy 1, $U^{2}AD$ achieved F1 scores of 0.7411 for patient-level and 0.5770 for segment-level on RenjiDCM T2w datset, which were lower than those of DAE and other models. However, after adapting on clinical data, $U^{2}AD$ showed significant improvements across all metrics.

\Cref{UAD_anomaly_comparison} illustrates two examples from RenjiDCM T2w dataset. We visualize their reconstruction results and anomaly maps for UAD models trained with their respective optimal strategies. For models utilizing Strategy 1, including DAE, f-AnoGAN, and AnoDDPM, the reconstruction performance was adversely affected by the domain shift between the training and test sets. For example, f-AnoGAN failed to reconstruct test images entirely. While DAE's anomaly maps successfully highlighted abnormal regions, high reconstruction errors in normal regions degraded its overall anomaly detection performance. Similarly, AnoDDPM struggled to reconstruct normal regions, impairing its ability to distinguish anomalies effectively.
For VAEGAN, which employed Strategy 2, overfitting to the anomaly regions was evident. This overfitting led to higher reconstruction errors in normal regions and lower errors in abnormal regions. As shown by the red arrows in the reconstructed images, hyperintensities were inaccurately recovered, resulting in detection failures.
Using Strategy 3, VAE improved its reconstruction performance for normal regions in test images. However, it partially recovered anomaly regions, limiting its anomaly detection effectiveness. In contrast, $U^{2}AD$ with Strategy 3 achieved an optimal balance between reconstructing normal regions and detecting anomalies. Its anomaly maps effectively highlighted reconstruction errors in abnormal regions, enabling clear distinction and accurate detection of hyperintensities. 

\Cref{UAD_anomaly_comparison} also depicts anomaly maps at various adaptation points for DAE, VAE, and $U^{2}AD$. All three models adopt the adaptation training process (Strategy 3) for the fair comparison. In the Example 1 and 2, the reconstruction error (annotated by yellow numbers in the anomaly maps) decreased in DAE and VAE during the early stages of adaptation, reflecting model adjustment to the test images and an improvement in overall reconstruction performance. However, by epoch 100, DAE began highlighting anomalies on the map but subsequently overfitted on the anomaly region, as evidenced by reduced reconstruction errors for anomaly regions (epoch 150) and increased errors in normal regions (epoch 200). Similarly, VAE exhibited overfitting in anomaly regions after epoch 150. In contrast, $U^{2}AD$ effectively avoids overfitting and consistently highlights anomalies throughout adaptation. 

\Cref{openneuro_signaldist} illustrates the signal intensity distributions of reconstructed pseudo-anomaly regions in the OpenNeuro T2w dataset. Specifically, the signal distribution of pseudo-anomaly regions in reconstructed images (orange) is compared against the normal SC signal (blue) and the pseudo-anomaly signal (red) from the original images. The goal of a UAD model is to reconstruct normal signals while highlighting the reconstruction error in pseudo-anomaly regions, such that the reconstructed pseudo-anomaly signals align closely with the normal SC signal and deviate from the original anomaly signal. For a fair comparison, all models were trained using Strategy 3 on the dataset. The results reveal that DAE and AnoDDPM fail to effectively capture the normal SC signal. In contrast, VAE and VAEGAN exhibit reconstructed distributions that fall between normal and anomaly signals. f-AnoGAN and $U^{2}AD$ achieve distributions closer to the normal SC regions, with $U^{2}AD$ demonstrating a shape highly similar to the normal signal distribution. However, the distribution of f-AnoGAN exhibits deviations at both ends, exceeding the normal signal range, which could introduce noise in anomaly maps. Further examples illustrating this issue are provided in the \ref{Reconstruction_analysis_on_pseudo-anomalies}.

\subsection{Comparison Experiments with Other Methods}
\Cref{Comparison of different methods} presents the anomaly detection results for various methods, including object detection-based models, signal analysis-based models, and our proposed $U^{2}AD$. The object detection-based models employ supervised learning, while the signal analysis-based models rely on statistical or clustering techniques.

On the RenjiDCM T2w dataset, $U^{2}AD$ demonstrated superior performance in the patient-level identification task, achieving the highest accuracy (0.7740), F1 score (0.8336), and specificity (0.5533). These results significantly outperformed the three supervised object detection-based models, which achieved F1 scores of 0.7545, 0.7540, and 0.7800 (for YOLOv8, Mask R-CNN, and DETR, respectively). Although these supervised models outperformed the signal analysis-based methods, they were still far below the performance of $U^{2}AD$.
In the segment-level localization task on RenjiDCM T2w, $U^{2}AD$ again ranked the highest, with an F1 score of 0.7345 and a recall of 0.8829. DETR achieved a comparable F1 score of 0.7332, ranking second, while YOLOv8 achieved an F1 score of 0.7103, outperforming our baseline method. The signal analysis-based methods performed the worst overall, with SLIC achieving the highest specificity (0.7750) but demonstrating the lowest F1 score and recall among all models. This highlights the limitations of these methods in handling complex clinical anomaly detection tasks.

On the RuijinDCM T2w dataset, $U^{2}AD$ clearly outperformed all other methods. In the patient-level identification task, $U^{2}AD$ achieved the highest accuracy (0.7740), F1 score (0.5569), and specificity (0.8556). The scarcity of anomalous data and associated annotations in RuijinDCM T2w posed a significant challenge for the supervised object detection-based methods. This lack of sufficient labeled anomalies resulted in poor generalization performance, as evidenced by the low F1 scores and specificity values of Mask R-CNN and DETR, which failed to achieve meaningful specificity. Signal analysis-based methods such as SLIC and DBSCAN performed better than object detection-based models but were still unable to match the performance of $U^{2}AD$.
In the segment-level localization task on RuijinDCM T2w, $U^{2}AD$ achieved the highest F1 score (0.6846), recall (0.7000), and specificity (0.9857). In contrast, YOLOv8, which performed relatively well on RenjiDCM T2w, delivered subpar results on RuijinDCM T2w. Signal analysis-based methods, such as SLIC and DBSCAN, showed reasonable specificity but struggled with low recall and F1 scores.

\begin{table*}[!ht]
\centering
\caption{Ablation for Monte-Carlo inference, EU-guided masking, and AU-guided masking.}
\label{Ablation}
\scalebox{0.45}{
\begin{tabular}{|l|c|c|c|c|c|c|c|c|c|c|c|c|c|}
\hline
\multicolumn{14}{|c|}{\textbf{on RenjiDCM T2w dataset}}\\ \hline
\multirow{2}{*}{\textbf{Model}} & \multicolumn{3}{c|}{\textbf{Modules}} & \multicolumn{3}{c|}{Patient-level identification} & \multicolumn{3}{c|}{Segment-level localization} & \multicolumn{4}{c|}{Reconstruction}\\
\cline{2-14}
& \textbf{MC} & \textbf{EG} & \textbf{AG} & Accuracy$\uparrow$ & F1 score$\uparrow$ & Specificity$\uparrow$ & F1 score$\uparrow$ & Recall$\uparrow$ & Specificity$\uparrow$ & SSIM$\uparrow$ & PSNR$\uparrow$ & MSE$\downarrow$ & Variance$\downarrow$ \\
\hline
MAE & $\times$ & $\times$ & $\times$ & 0.6962 ± 0.0275 & 0.7863 ± 0.0427 & 0.3625 ± 0.1386 & 0.7160 ± 0.0869 & 0.7935 ± 0.1440 & 0.6648 ± 0.1316 & / & / & / & / \\
baseline & \checkmark & $\times$ & $\times$ & 0.6795 ± 0.1269 & 0.8127 ± 0.0425 & 0.4918 ± 0.1124 & 0.7049 ± 0.0705 & 0.7899 ± 0.1166 & \textbf{0.6885 ± 0.1352} & 0.7472  & 21.0376 & 0.0048 & 2.2200 \\
EU-guided & \checkmark & \checkmark & $\times$ & 0.7853 ± 0.1569 & 0.8170 ± 0.0397 & 0.5014 ± 0.1359 & 0.7259 ± 0.0795 & 0.8332 ± 0.1354 & 0.5383 ± 0.1811 & 0.7600  & 21.5713 & 0.0043 & 1.7201\\
$U^{2}AD$ & \checkmark & \checkmark & \checkmark & \textbf{0.8537 ± 0.0763} & \textbf{0.8336 ± 0.0348} & \textbf{0.5533 ± 0.1384} & \textbf{0.7345 ± 0.0547} & \textbf{0.8829 ± 0.0977} & 0.5299 ± 0.1379 & \textbf{0.7637} & \textbf{21.6574} & \textbf{0.0042} & \textbf{1.7343} \\
\hline
\multicolumn{14}{|c|}{\textbf{on RuijinDCM T2w dataset}}\\ \hline
\multirow{2}{*}{\textbf{Model}} & \multicolumn{3}{c|}{\textbf{Modules}} & \multicolumn{3}{c|}{Patient-level identification} & \multicolumn{3}{c|}{Segment-level localization} & \multicolumn{4}{c|}{Reconstruction}\\
\cline{2-14}
& \textbf{MC} & \textbf{EG} & \textbf{AG} & Accuracy$\uparrow$ & F1 score$\uparrow$ & Specificity$\uparrow$ & F1 score$\uparrow$ & Recall$\uparrow$ & Specificity$\uparrow$ & SSIM$\uparrow$ & PSNR$\uparrow$ & MSE$\downarrow$ & Variance$\downarrow$ \\
\hline
MAE & $\times$ & $\times$ & $\times$ & 0.5363 ± 0.1918 & 0.3100 ± 0.1537 & 0.5033 ± 0.2905 & 0.6073 ± 0.3701 & 0.6600 ± 0.4025 & 0.9761 ± 0.0343 & / & / & / & /\\
baseline & \checkmark & $\times$ & $\times$ & 0.7440 ± 0.0166 & 0.3856 ± 0.1282 & 0.6574 ± 0.2098 & 0.5896 ± 0.3749 & 0.6050 ± 0.4031 & \textbf{0.9860 ± 0.0241} & 0.7762 & 19.9671 & 0.0074 & 2.9136 \\
EU-guided & \checkmark & \checkmark & $\times$ & 0.7516 ± 0.0099 & 0.5116 ± 0.1591 & 0.7770 ± 0.2497 & 0.6606 ± 0.3401 & 0.6800 ± 0.3702 & 0.9787 ± 0.0288 & 0.7891 & 20.4865 & 0.0067 & 2.2446 \\
$U^{2}AD$ & \checkmark & \checkmark & \checkmark & \textbf{0.7740 ± 0.0124} & \textbf{0.5569 ± 0.2333} & \textbf{0.8556 ± 0.1577} & \textbf{0.6846 ± 0.3384} & \textbf{0.7000 ± 0.3666} & 0.9857 ± 0.0270 & \textbf{0.7904} & \textbf{20.5160} & \textbf{0.0066} & \textbf{2.2274} \\
\hline
\multicolumn{10}{l}{\textbf{Bold} values indicate the best performance for each metric.}  \\
\multicolumn{14}{l}{MC: Monte-Carlo inference for reconstruction; EG: EU-guided masking; AG: AU-guided masking} \\
\multicolumn{10}{l}{$\uparrow$ indicates the higher value is better, and $\downarrow$ indicates the lower is better.} 
\end{tabular}
}
\end{table*}

\subsection{Ablation Studies}
As shown in \Cref{Ablation}, we conducted ablation studies on two datasets to assess the effectiveness of each module in the proposed $U^{2}AD$ framework. Specifically, MAE and baseline models employed the traditional random masking strategy during adaptation training, while the EU-guided model used a one-stage EU-guided masking strategy, and $U^{2}AD$ adopted the proposed two-stage EU/AU-guided masking approach. In terms of reconstruction, the MAE model performed single-pass reconstruction for each image patch, whereas baseline, EU-guided, and $U^{2}AD$ utilized Monte Carlo inference to reconstruct each patch $K = 10$ times. The results demonstrated that Monte Carlo inference significantly improved anomaly detection performance on both datasets, highlighting its importance in achieving robust reconstructions. Furthermore, the EU-guided masking strategy improved F1 scores for both patient-level and segment-level anomaly detection tasks compared to the baseline, while the two-stage EU/AU-guided masking in $U^{2}AD$ delivered the best overall performance. Furthermore, $U^{2}AD$ consistently achieved the highest SSIM and RSNR as well as the lowest MSE and Variance on both RenjiDCM T2w and RuijinDCM T2w datasets, underscoring the effectiveness of integrating uncertainty-guided masking strategies for superior performance. 

\subsection{Sensitivity Studies}
%This section includes three parts: hyperparameter experiments, robustness experiments, and stability experiments.
\subsubsection{Hyperparameter Selection}
In our proposed model, there are three key hyperparameters: (1) the number of Monte Carlo inference, denoted as \(K\), (2) the temperature coefficient $\tau$ used to control patch probability in the EU-guided adaptation, and (3) the mask ratio (r) which indicates the ratio of masked patches in all ROI patchs. To investigate the impact of these parameters on our model's performance, we conducted experiments with different hyperparameter values on RenjiDCM T2w dataset.

%The effects of different \(K\) values on image reconstruction and anomaly detection results are shown in \Cref{hyperparameters}. In \Cref{hyperparameters} (a), we observe that the reconstruction results of our model are not sensitive to \(K\). When \(K\) increases from 3 to 5 and then to 10, SSIM and PSNR show slight improvements, while MSE decreases slightly. However, no further improvements are observed when $K$ is increased beyond 10 (i.e., for $K = 10$, $K = 15$, and $K = 20$). 
\Cref{hyperparameter performance} shows the impact of different $K$ values on anomaly detection results. For $K = 10$, $K = 15$, and $K = 20$, the anomaly detection performance is consistently good, with patient-level F1 scores above 0.83 and segment-level F1 scores above 0.73. However, increasing $K$ means more sampling steps are required, which leads to a significant increase in inference time. Specifically, when $K$ increases from 3 to 10, the average inference time per image increases from 0.0639s to 0.1521s. When $K$ increases from 10 to 20, the average inference time increases from 0.1521s to 0.2743s. To achieve an optimal balance between model performance and inference time, we selected $K = 10$.

The temperature coefficient $\tau$ controls the influence of the EU value of each patch on its probability of being selected. When $\tau > 0$, a smaller $\tau$ leads to a higher probability for patches with larger EU values, and vice versa. 
%In \Cref{hyperparameters} (b), we show the impact of different $\tau$ values on the model's reconstruction results. It can be seen that $\tau$ does not significantly affect reconstruction performance in terms of SSIM, RSNR, and MSE. However, it is worth noting that when $\tau = 0$, the probability weights of different patches are uniform, effectively becoming random masking as in the baseline model. From \Cref{reconstruction_trend}, we can observe that models with $\tau > 0$ show significant improvements in reconstruction results compared to the baseline with $\tau = 0$.
\Cref{hyperparameter performance} shows the results of anomaly detection for different $\tau$ values. It is demonstrated that $\tau=1.0$ achieved the best performance among all values.

Higher mask ratio indicates that more patches are masked during training, with less visible information available. 
%\Cref{hyperparameters} (c) demonstrates that higher mask ratio leads to worse reconstruction within the masked regions. 
\Cref{hyperparameter performance} gives the results of anomaly detection tasks for different ratios. When ratio is too low or too high, its performance decreases. When ratio=0.75, the anomaly detection achieved the best results, with a relatively short inference time. Based on these results, ratio=0.75 strikes an optimal balance between performance and inference time.

\begin{table*}[!ht]
\centering
\caption{Performance metrics and average inference time for different hyperparameters of $U^{2}AD$. Experiments are performed on RenjiDCM T2w dataset.}
\label{hyperparameter performance}
\scalebox{0.60}{
\begin{tabular}{|c|c|c|c|c|c|c|c|c|}
\hline
\multicolumn{2}{|c|}{\multirow{2}{*}{\textbf{Hyperparameter}}} & \multicolumn{3}{c|}{Patient-level identification} & \multicolumn{3}{c|}{Segment-level localization} & \multirow{2}{*}{\parbox{1.5cm}{Average\\Time(s)}} \\ \cline{3-8} 
\multicolumn{2}{|c|}{} & Accuracy ↑ & F1 score ↑ & Specificity ↑ & F1 score ↑ & Recall ↑ & Specificity ↑ &  \\ \hline

\multirow{5}{*}{\parbox{1.5cm}{\(K\) values}}
                           & \(K\)=3    & 0.7500 ± 0.0153    & 0.8169 ± 0.0378    & 0.4954 ± 0.1058    & 0.6968 ± 0.0767    & 0.8188 ± 0.1396    & 0.4943 ± 0.1667    & 0.0639 \\
                           & \(K\)=5    & 0.7364 ± 0.0255    & 0.8090 ± 0.0385    & 0.4581 ± 0.1347    & 0.7133 ± 0.0709    & 0.7735 ± 0.1244    & 0.5279 ± 0.1541    & 0.0871 \\ 
                           & $K=10^{\star}$   & 0.7740 ± 0.0124    & 0.8336 ± 0.0348    & \textbf{0.5533 ± 0.1384}    & \textbf{0.7345 ± 0.0547}    & 0.8829 ± 0.0977    & 0.5299 ± 0.1379     & 0.1521 \\
                           & \(K\)=15   & \textbf{0.7744 ± 0.0131}    & \textbf{0.8348 ± 0.0390}    & 0.5355 ± 0.1037    & 0.7317 ± 0.0683    & 0.8045 ± 0.1334    & 0.5709 ± 0.1467    & 0.2131 \\
                           & \(K\)=20   & 0.7712 ± 0.0152    & 0.8323 ± 0.0382    & 0.5307 ± 0.0947    & 0.7343 ± 0.0601    & \textbf{0.9078 ± 0.1174}    & \textbf{0.5990 ± 0.1536}    & 0.2743 \\ 
                           \midrule[1.0pt]

\multirow{7}{*}{\parbox{1.5cm}{mask \\ratios (r)}}
                           & r=0.35 & 0.7272 ± 0.0245 & 0.8041 ± 0.0386 & 0.4309 ± 0.1613 & 0.7099 ± 0.0786 & 0.8125 ± 0.1241 & 0.6438 ± 0.1177 & 0.3901 \\ 
                           & r=0.45 & 0.7220 ± 0.0253 & 0.7999 ± 0.0401 & 0.4253 ± 0.1642 & 0.6923 ± 0.0807 & 0.7625 ± 0.1466 & 0.6633 ± 0.1546 & 0.2884 \\ 
                           & r=0.55 & 0.7304 ± 0.0216 & 0.8051 ± 0.0374 & 0.4446 ± 0.1304 & 0.7143 ± 0.0696 & 0.7962 ± 0.1243 & 0.6591 ± 0.1052 & 0.2238 \\
                           & r=0.65 & 0.7556 ± 0.0141 & 0.8198 ± 0.0339 & 0.5127 ± 0.1267 & 0.7137 ± 0.0795 & 0.8200 ± 0.1240 & 0.6243 ± 0.1321 & 0.1819 \\
                           & $r=0.75^{\star}$ & \textbf{0.7740 ± 0.0124} & \textbf{0.8336 ± 0.0348} & \textbf{0.5533 ± 0.1384} & \textbf{0.7345 ± 0.0547} & \textbf{0.8829 ± 0.0977} & 0.5299 ± 0.1379 & 0.1521 \\ 
                           & r=0.85 & 0.7492 ± 0.0156 & 0.8161 ± 0.0417 & 0.4953 ± 0.1057 & 0.7294 ± 0.0681 & 0.8742 ± 0.1229 & 0.5476 ± 0.1487 & 0.1231 \\ 
                           & r=0.95 & 0.7284 ± 0.0202 & 0.8039 ± 0.0419 & 0.4401 ± 0.1000 & 0.6974 ± 0.0743 & 0.7516 ± 0.1226 & \textbf{0.6810 ± 0.1438} & 0.0970 \\ 
                           \midrule[1.0pt]

\multirow{7}{*}{\parbox{1.5cm}{$\tau$ values}}
                           & $\tau$=0.1 & 0.7412 ± 0.0136 & 0.8113 ± 0.0416 & 0.4843 ± 0.0973 & 0.6928 ± 0.0702 & 0.7881 ± 0.1239 & 0.6192 ± 0.1419 & \multirow{7}{*}{/} \\ 
                           & $\tau$=0.2 & 0.7504 ± 0.0129 & 0.8181 ± 0.0418 & 0.4970 ± 0.0820 & 0.6992 ± 0.0864 & 0.7826 ± 0.1402 & \textbf{0.6523 ± 0.1209} &  \\ 
                           & $\tau$=0.5 & 0.7592 ± 0.0212 & 0.8201 ± 0.0504 & 0.5377 ± 0.1390 & 0.7046 ± 0.0780 & 0.8124 ± 0.1411 & 0.6004 ± 0.1747 &  \\ 
                           & $\tau=1.0^{\star}$ & \textbf{0.7740 ± 0.0124} & \textbf{0.8336 ± 0.0348} & \textbf{0.5533 ± 0.1384} & \textbf{0.7345 ± 0.0547} & \textbf{0.8829 ± 0.0977} & 0.5299 ± 0.1379 &  \\
                           & $\tau$=2.0 & 0.7576 ± 0.0156 & 0.8198 ± 0.0559 & 0.5221 ± 0.0976 & 0.6998 ± 0.0637 & 0.8316 ± 0.1301 & 0.5608 ± 0.1552 &  \\ 
                           & $\tau$=5.0 & 0.7520 ± 0.0160 & 0.8191 ± 0.0356 & 0.4976 ± 0.1125 & 0.7089 ± 0.0766 & 0.7850 ± 0.1279 & 0.6468 ± 0.1601 &  \\ 
                           & $\tau$=10.0& 0.7608 ± 0.0164 & 0.8239 ± 0.0385 & 0.5243 ± 0.1030 & 0.6948 ± 0.0742 & 0.8410 ± 0.1453 & 0.5287 ± 0.1303 &  \\ \hline
\multicolumn{8}{l}{$\star$: the optimal value of each hyperparameter for $U^{2}AD$.} \\
\multicolumn{8}{l}{\textbf{Bold} values indicate the best performance for each metric and hyperparameter.} \\
\multicolumn{8}{l}{↑ indicates the higher value is better.} 
\end{tabular}
}
\end{table*}

\begin{figure*}[!ht]
\centerline{\includegraphics[width=\columnwidth]{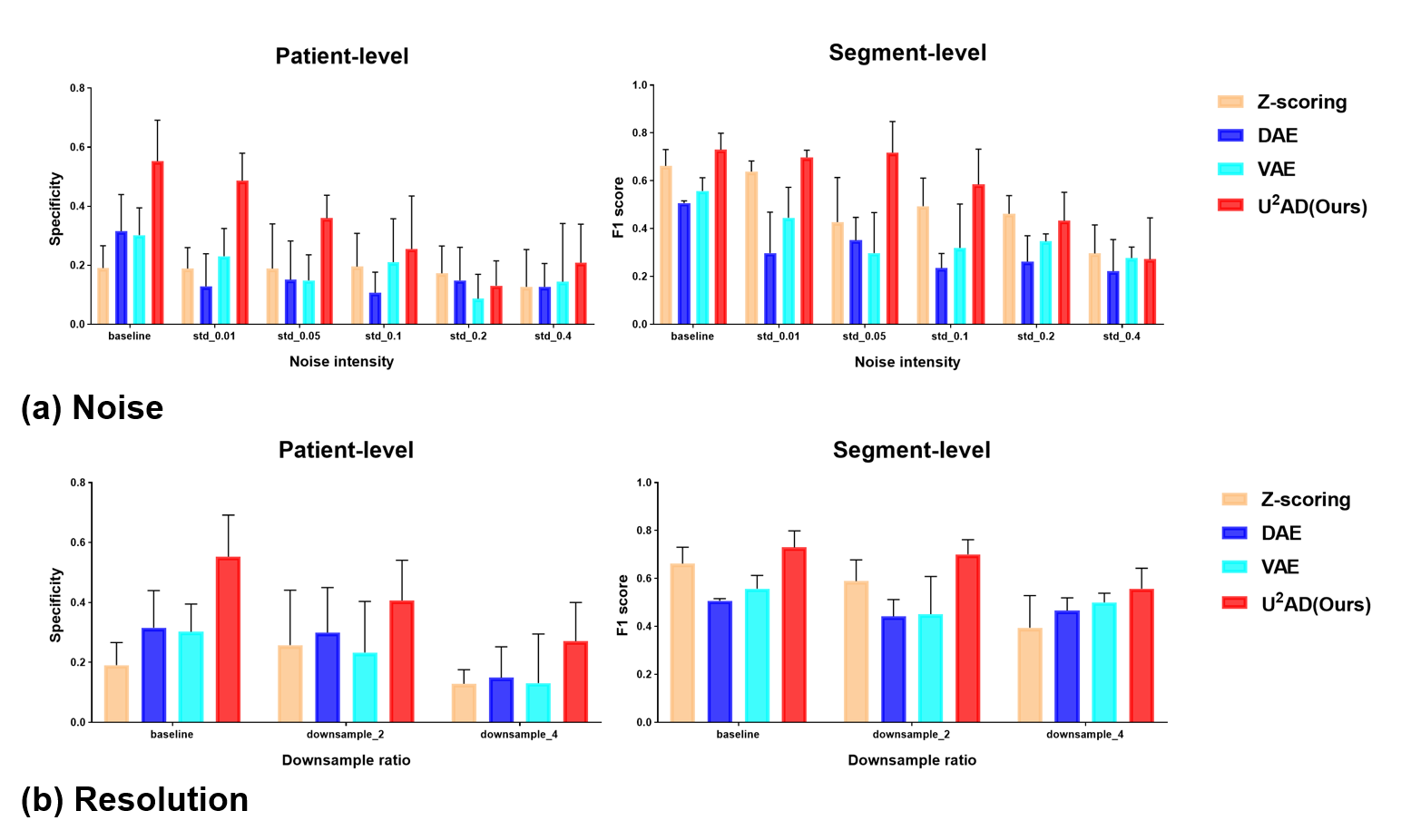}}
\caption{Robustness experiments evaluating the impact of noise and image resolution on the proposed method. (a) Performance under varying levels of Gaussian noise, where standard deviations ($\sigma$) are used. (b) Performance under image downsampling, where the original images are downsampled by factors of 2 and 4.}
\label{robustness}
\end{figure*}

\subsubsection{Robustness to Image Noise and Resolution}
To evaluate the robustness of our model to noise and image resolution under different conditions, we added varying levels of Gaussian noise to the images and performed downsampling to assess reconstruction and anomaly detection. The experiments are performed on RenjiDCM T2w dataset. We compared our model with three other unsupervised methods: Z-Scoring, DAE, and VAE. The Gaussian noise added had a mean of 0 and variances ranging from 0 to 0.4. The results are shown in \Cref{robustness}(a). All unsupervised models showed performance degradation after noise addition, particularly in the patient-level detection task. When the noise variance exceeded 0.2, the specificity of all models dropped below 0.2.

$U^{2}AD$ shows good robustness to noise in segment-level localization. Even with Gaussian noise of variance 0.05, $U^{2}AD$'s F1 score remained above 0.7, whereas Z-Scoring, DAE, and VAE saw significant drops in F1 scores. When the noise variance exceeded 0.1, the performance of all models dropped significantly. However, since real-world images are unlikely to have very high noise levels, $U^{2}AD$ retains good robustness when the noise variance is below 0.01, outperforming the comparative models.

For the resolution experiments, we performed downsampling of the original images by factors of 2 and 4. The results are shown in \Cref{robustness}(b). As the downsampling factor increased, model performance in both patient-level and segment-level tasks declined. Notably, the performance decline in the patient-level task was more pronounced than in the segment-level task. Among the models, $U^{2}AD$ consistently delivered the best anomaly detection results for images with different resolutions.

%This indicates that anomaly detection in low-resolution images, especially in identifying patients with abnormalities, is significantly affected. Among the models, our method consistently delivered the best anomaly detection results for images with different resolutions.

\section{Discussion}
%In this study, we presented the $U^{2}AD$ framework, a novel uncertainty-guided UAD approach specifically designed for detecting T2 hyperintensities in spinal cord MRI. Our experimental results demonstrate that $U^{2}AD$ consistently outperforms state-of-the-art UAD methods and supervised object detection models in both patient-level anomaly identification and segment-level localization tasks. By incorporating aleatoric and epistemic uncertainty into TTA, $U^{2}AD$ effectively addresses critical challenges in anomaly detection, including domain shifts and the inherent task conflict between reconstruction objectives and anomaly detection. We further evaluated three distinct dataset training strategies, under the assumption of a domain shift between healthy and clinical datasets. The results highlight that $U^{2}AD$ achieves the most significant improvements when employing a two-step training strategy: pretraining on a large healthy dataset followed by adaptation to the target clinical dataset. Moreover, hyperparameter tuning, robustness testing, and reliability experiments confirm the robustness and adaptability of the proposed $U^{2}AD$ framework across varying conditions and datasets.
\subsection{Performance analysis of different methods}
%Compared to other UAD methods, including VAE, DAE, and AnoDDPM, $U^{2}AD$ demonstrates superior performance in distinguishing anomalies from normal regions. Importantly, for the patient-level task, we selected an anomaly score threshold corresponding to a recall of 0.90, ensuring high sensitivity for anomaly detection in the validation set. While the high recall necessitates a relatively low threshold, which may reduce specificity, this configuration aligns with clinical requirements where minimizing false negatives is critical. Despite this tradeoff, $U^{2}AD$ achieved the highest specificity of 0.5533 and 0.8556 in RenjiDCM T2w and Ruijin T2w dataset, respectively, significantly surpassing other models, most of which fell below 0.35. In the segment-level localization task, the goal is to assess the model's ability to accurately identify anomalies within specific vertebral segments. Here, the anomaly score threshold was determined based on the F1 score that maximized performance in the training set. Although $U^{2}AD$ did not achieve the best recall or specificity individually in RenjiDCM T2w dataset, it effectively balanced these metrics, ensuring robust and reliable performance in clinical applications.

The proposed $U^{2}AD$ framework outperformed other UAD methods, such as VAE, DAE, and AnoDDPM, in distinguishing anomalies from normal regions. For the patient-level task, a recall-oriented threshold (0.90) ensured high sensitivity, meeting clinical needs where minimizing false negatives is crucial. Despite this sensitivity-focused configuration, $U^{2}AD$ achieved the highest specificity in both RenjiDCM T2w and Ruijin T2w datasets, significantly exceeding competing models. In segment-level localization, $U^{2}AD$ effectively balanced recall and specificity, offering robust and reliable performance suited to clinical applications.

Object detection models, including YOLOv8, Mask R-CNN, and DETR, showed competitive performance in localization tasks but struggled with identification, particularly in small and annotation-sparse datasets like RuijinDCM T2w. Traditional signal analysis methods, while training-free, underperformed due to sensitivity to signal noise and dependence on manual hyperparameter tuning, further emphasizing the advantages of $U^{2}AD$ in handling complex clinical scenarios.

%这里讨论的关键点是提出了一种新的数据集训练策略，使得无监督异常检测在临床上更加实用，并且和其他两个策略和其他UAD方法进行了全面对比，展现了不同策略各自存在的问题。在本任务中，预训练＋测试时适应结合我们的模型可以取得一个最佳的效果。
\subsection{Analysis of different dataset training strategies}
Currently most UAD researches predominantly conduct experiments on public datasets, such as the BraTS2021 dataset \citep{baid2021rsna}, RSNA \citep{shih2019augmenting}, and VinDr-CXR \citep{nguyen2022vindr}. These datasets are typically pre-divided into healthy images for training and potentially anomalous images for testing. While effective for benchmarking, this setup is often impractical in real-world clinical settings, where such manual partitioning is neither feasible nor representative of clinical data diversity.
It is challenging to simultaneously obtain large datasets of healthy and clinical data from a single distribution. Even if such a dataset exists, manually partitioning healthy training samples from anomalous test samples renders the anomaly detection task ``semi-supervised''. In practice, healthy datasets are relatively easy to obtain and are often large in scale, whereas target datasets requiring anomaly detection are typically small and collected from clinical centers, often with significantly different signal distributions compared to the healthy dataset. This discrepancy is evident in our study, where we utilized a large-scale healthy public dataset and clinical datasets collected from different sites, with distinct signal distributions as shown in \Cref{introduction}. To address these challenges, this study compared three dataset training strategies for different UAD models. %Strategy 1 involves training exclusively on the healthy public dataset and evaluating on the clinical dataset. Strategy 2 trains and evaluates solely on the clinical dataset without utilizing healthy data. Strategy 3 adopts a two-step approach: pretraining on the healthy dataset followed by test-time adaptation on the clinical dataset. 
Each strategy presents unique advantages and limitations. In Strategy 1, the model learns a robust representation of healthy images but may fail to reconstruct test images accurately due to domain shifts. Strategy 2 allows the model to fit the clinical dataset better but risks learning representations of anomalies, which contradicts the core objective of anomaly detection. Strategy 3 aims to combine the strengths of both approaches. Pretraining on the large healthy dataset initializes the model with strong representation capabilities for normal image reconstruction, while adaptation training learns the target clinical dataset’s distribution. However, the adaptation process may lead to overfitting on anomalous regions. Overall, Strategy 3 offers a potential solution by balancing these trade-offs and demonstrating superior performance combined with the proposed $U^{2}AD$, as illustrated in \Cref{UAD_anomaly_comparison} and \Cref{Comparison of UAD Methods}. In the future, further experiments and analysis in different UAD tasks and datasets on these three strategies are needed.

\subsection{Uncertainty application in solving task conflict of UAD}
Many studies have highlighted the inherent conflict between reconstruction in normal and anomalous regions in anomaly detection tasks. \citet{cai2024rethinking} argued that relying on autoencoders trained exclusively on normal data to enable anomaly detection based on reconstruction errors does not always hold, due to a mismatch between the reconstruction objective and the anomaly detection task objective. They identified an ``identical shortcut'' in reconstruction networks, which allows abnormal regions to be reconstructed successfully, leading to false negatives. Similarly, \citet{liang2024itermask} observed that the prevalent ``corrupt-and-reconstruct'' paradigm in reconstruction-based UAD models often results in suboptimal reconstruction of even normal regions, causing false positives. This creates a ``paradox'': the ideal scenario for anomaly detection occurs when the model achieves overfitting in normal regions and underfitting in anomaly regions. Traditional UAD models often struggle to strike an optimal balance between these two objectives. \Cref{UAD_anomaly_comparison} illustrates this challenge during the training process of DAE and VAE. To address this issue, \citet{shvetsova2021anomaly} proposed transforming the fully unsupervised training process into a weakly supervised approach, incorporating a validation set with a small number of abnormal images. While their results indicated that this strategy aids in determining hyperparameters and stop points for training, individual differences between validation and target images often render the optimal stop point suboptimal for unseen test data. 

%\Cref{UAD_cases} illustrates anomaly maps at various adaptation points for DAE, VAE, and $U^{2}AD$. All three models were pretrained on a healthy dataset and subsequently adapted to the clinical dataset. In Case 1, the reconstruction error (annotated by yellow numbers in the anomaly maps) decreased in DAE and VAE during the early stages of adaptation, reflecting model adjustment to the test images and an improvement in overall reconstruction performance. However, by epoch 100, DAE began highlighting anomalies on the map but subsequently overfitted on the anomaly region, as evidenced by reduced reconstruction errors for anomaly regions (epoch 150) and increased errors in normal regions (epoch 200). Similarly, VAE exhibited overfitting in anomaly regions after epoch 150. To address this issue, Shvetsova et al.\cite{shvetsova2021anomaly} proposed transforming the fully unsupervised training process into a weakly supervised approach, incorporating a validation set with a small number of abnormal images. While their results indicated that this strategy aids in determining hyperparameters and stop points for training, individual differences between validation and target images often render the optimal stop point suboptimal for unseen test data. 

While some recent researches \citep{cai2024rethinking,liang2024itermask} have made progress in addressing reconstruction conflicts by utilizing information theory or iterative masking strategy, there has been limited exploration of leveraging uncertainty during reconstruction training in UAD models. In our study, we not only use uncertainty as an indicator of potential anomalies but also integrate it directly into model optimization. The variance map (EU) derived from these reconstructions reflects the model’s confidence and is referred to as epistemic uncertainty. EU guides the masking and reconstruction training, optimizing reconstruction performance in normal regions, particularly those with complex anatomical structures. Meanwhile, the error map (AU) between the average reconstruction and the original image represents aleatoric uncertainty, which highlights potential anomalies. High-AU regions are selected as anomaly candidates and excluded from training, ensuring underfitting and high reconstruction errors in anomaly regions. This dual uncertainty approach reduces false negatives by maintaining high reconstruction errors in anomaly regions and minimizes false positives by improving reconstruction accuracy in normal regions. %For Case 1 in \Cref{UAD_cases}, $U^{2}AD$ effectively avoids overfitting and consistently highlights anomalies throughout adaptation.

\begin{figure*}[!t]
\centerline{\includegraphics[width=\columnwidth]{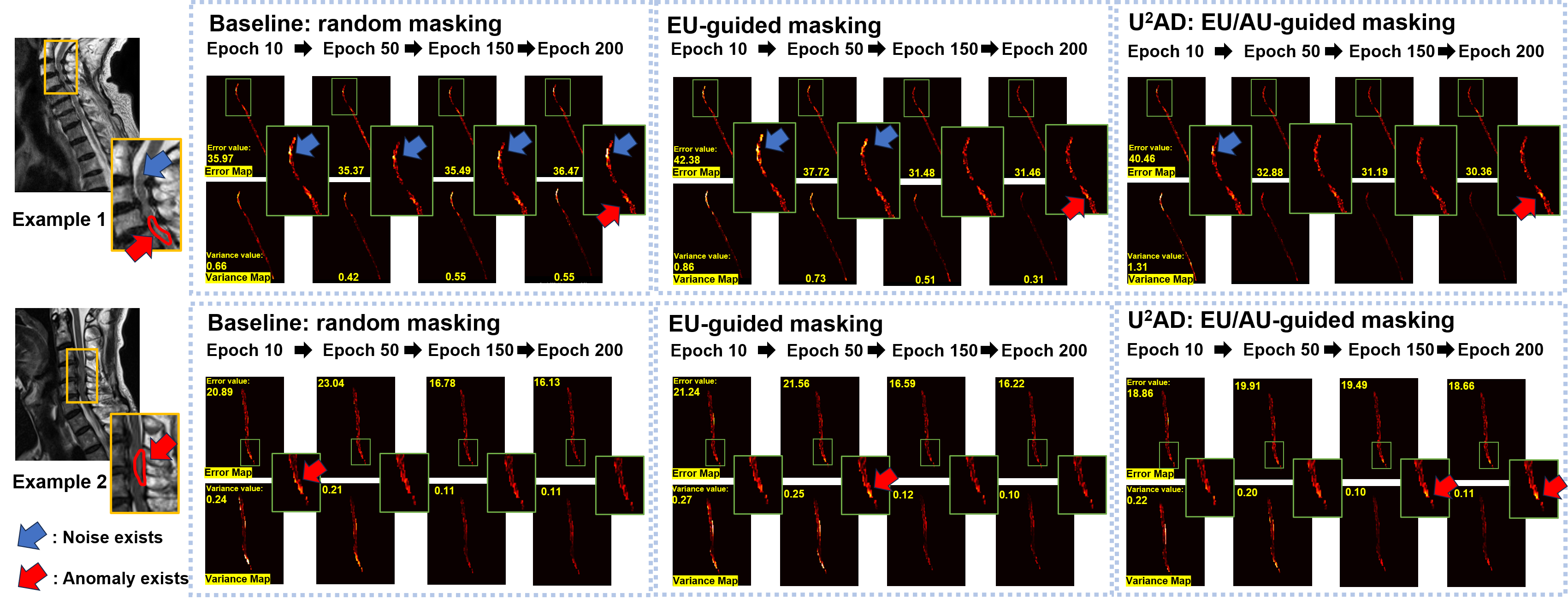}}
\caption{Visualization of Error maps (aleatoric uncertainty, AU) and Variance maps (epistemic uncertainty, EU) at different epochs during adaptation training. The figure compares the baseline model (random masking), EU-guided model, and the proposed $U^{2}AD$ framework to evaluate the effectiveness of the masking strategies in distinguishing anomaly regions. Yellow boxes highlight the ROIs, while blue arrows indicate the noises, and red arrows mark the real anomaly regions. The error and variance values are displayed in the each map for quantitative assessment. }
\label{U2AD_examples}
\end{figure*}

\subsection{Analysis of different masking strategies}
Notably, our baseline model utilized a random masking strategy during adaptation, while $U^{2}AD$ incorporated the proposed UG-Ada, which leverages both EU-guided masking and AU-guided masking strategies. \Cref{Ablation} has proven the effectiveness of proposed uncertainty-guided masking strategies quantitatively. \Cref{U2AD_examples} further compares the performance of these three masking strategies by illustrating error maps and variance maps at various adaptation epochs.
In Example 1, the baseline model exhibited high variance and elevated reconstruction errors in normal regions, as indicated by the blue arrows. These false positives occurred due to insufficient training in low-confidence regions. By introducing the EU-guided masking strategy, $U^{2}AD$ directed prioritized training to these regions, thereby reducing variance and reconstruction errors and mitigating false positives.
In Example 2, the baseline and EU-guided models successfully detected anomaly regions at epochs 10 and 50, respectively. However, as adaptation progressed, the reconstruction error within the anomaly regions decreased, causing these anomalies to blend into the surrounding regions and resulting in false negatives. To address this limitation, $U^{2}AD$ incorporated the AU-guided masking strategy, which identifies high-AU (high-error) patches as anomaly candidates and excludes them from training. By preventing these patches from being reconstructed during adaptation, $U^{2}AD$ maintained elevated reconstruction errors in anomaly regions, thereby enhancing their distinction and improving overall detection accuracy.

\subsection{Clinical applications for analysing T2 hyperintensity lesions}
In the context of DCM, T2 hyperintensity serves as an important indicator of various pathological changes within the spinal cord, including edema, inflammation, hemorrhage, ischemia, necrosis, and gliosis, providing crucial information for disease diagnosis \citep{nouri2016magnetic}. Additionally, it is closely associated with surgical prognosis. Studies such as those by \citet{yukawa2007mr} have shown that different grades of T2 hyperintensity may have a poorer postoperative JOA score improvement rate, which can predict surgical outcomes. The signal change ratio (SCR), a quantitative measure, is related to upper and lower limb neurological impairment, with higher SCR values often indicating more severe impairment \citep{nouri2017relationship}.

However, there are notable challenges in clinical applications. The lack of a unified ``gold standard'' for T2 hyperintensity evaluation is a major issue. Qualitative methods, such as the simple dichotomous method, fail to provide detailed lesion information. The three - level classification method is influenced by MRI image quality and the subjectivity of evaluating physicians. Quantitative methods like SCR calculation are time - consuming, rely on manual operation, and are prone to errors. Moreover, measuring the area of T2 hyperintensity is difficult due to significant variations across different image planes and high dependence on professional software \citep{wei2020does}. Additionally, the application of artificial intelligence in evaluating T2 hyperintensity in DCM is still in its infancy, unable to meet the clinical demand for efficient and accurate assessments. The framework developed in this study can accurately identify and locate lesions, laying a solid foundation for subsequent quantitative and intelligent analysis of T2 hyperintensity lesions.

\subsection{Limitations and future work}
This study has several limitations. First, our method relies on Monte Carlo estimation of uncertainty, where the inference time increases with the number of inference iterations (\(K\)). While a larger \(K\) improves model performance and reliability, as demonstrated in \Cref{hyperparameter performance}, it also raises computational demands. To balance performance and efficiency, we selected an optimal \(K\) value of 10, resulting in an inference time of 0.1521 seconds per image, which remains acceptable for clinical applications.
Second, this study focuses solely on the application of $U^{2}AD$ to the detection of T2 hyperintensities in spinal cord MR images. Future work will explore the generalizability of $U^{2}AD$ by validating its performance across other anomaly detection tasks and datasets in diverse clinical contexts.
Third, the current implementation of $U^{2}AD$ is designed for 2D images, while many medical imaging modalities, such as CT and MRI, provide three-dimensional data. In future research, we aim to extend $U^{2}AD$ to 3D images, enabling its application to a broader range of medical imaging tasks and further enhancing its clinical utility.

\section{Conclusion}
In this paper, we proposed $U^{2}AD$, an uncertainty-based unsupervised anomaly detection framework designed to address the critical challenges of domain shifts and task conflicts in clinical applications. To enhance its performance on clinical datasets, we introduced a two-step pretraining and adaptation strategy, which effectively leverages large-scale healthy public datasets for initialization and adapts to domain-specific clinical data. By combining epistemic uncertainty and aleatoric uncertainty, the framework guides the adaptation process to address training imbalances between normal and abnormal regions. This enables $U^{2}AD$ to achieve high reconstruction fidelity in normal regions while amplifying errors in abnormal areas, effectively improving anomaly detection sensitivity. When applied to the detection task of T2 hyperintensities in spinal cord MR images, a key biomarker for conditions such as degenerative cervical myelopathy, $U^{2}AD$ outperformed state-of-the-art UAD methods and supervised approaches, demonstrating its robust and adaptable design. Furthermore, the framework’s flexible architecture underscores its potential for broader applications in various anomaly detection tasks.

\section{Author contributions}
Q.Z., X.Y.C., Z.Y.H., and K.W. contributed equally to this work and should be regarded as co-first authors. Q.Z., X.Y.C, and Z.Y.H. contributed the clinical design, experimental design, data processing, data annotation, method validation, literature research and manuscript writing. L.M.W. and K.W. participated in the calculation of clinical indicators, model comparison, and providing guidance for clinical applications. K.W. was also involved in tasks related to data annotation, data analysis and processing. J.Q.S. participated in the completion of theoretical research, experimental design, and manuscript revision. H.X.S. participated in the completion of theoretical research, experimental design, clinical application guidance and manuscript revision. 

\section{Declaration of interests}
Our work has no financial interest.

\section{Funding sources}
This research was supported by the National Natural Science Foundation of China (Grant No.62471297 and No.82272476), 2022 Shanghai Leading Project of the Oriental Talents Program, the National Key Research and Development Program (Grant No.2023YFC2411401), Shanghai Jiao Tong University Medical Engineering Cross Research Funds (Grant No.YG2021ZD05), Shanghai Zhangjiang National Independent Innovation Demonstration Zone Special Development Fund Major Project (Grant No.ZJ2021-ZD-007), and Clinical Postdoctoral Program of Shanghai Jiao Tong University School of Medicine and Talent Cultivation Project of Renji Hospital (Grant No.RJTJ25-QN-031).

\appendix

\section{Experiments on the OpenNeuro T2w}
\setcounter{figure}{0}
\renewcommand{\thefigure}{A\arabic{figure}}
\begin{figure*}[!ht]
\centerline{\includegraphics[width=\columnwidth]{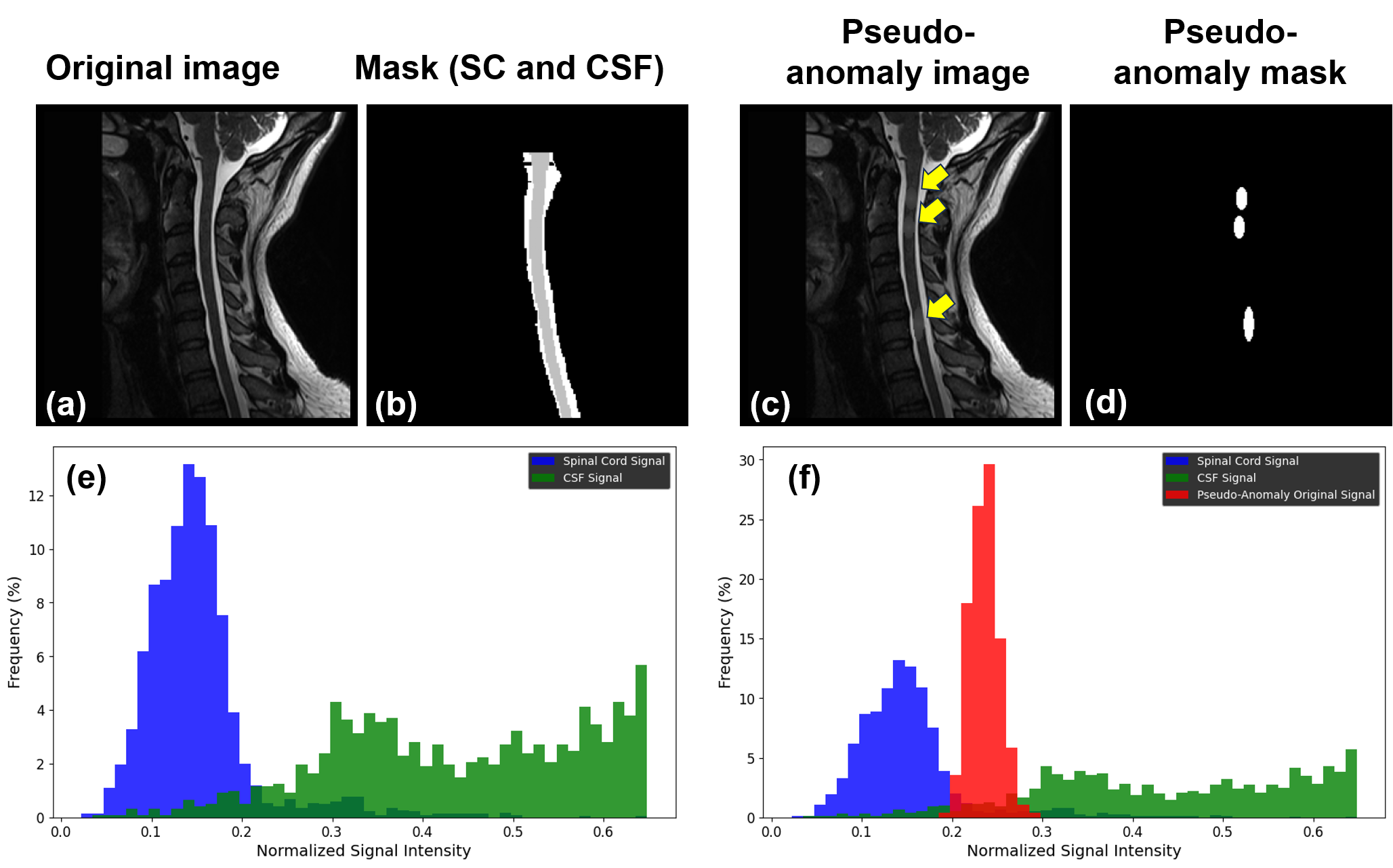}}
\caption{An example of adding pseudo-anomalies into normal (healthy) ROI regions. (a) Original MR image; (b) corresponding SC (gray) and CSF (white) masks; (e) normalized signal intensity distribution of SC and CSF; (d) generated pseudo-anomaly mask; (f) pseudo-anomaly signal distribution; (c) final pseudo-anomaly image.}
\label{pseudo-anomaly}
\end{figure*}

\renewcommand{\thefigure}{A\arabic{figure}}
\begin{figure*}[!ht]
\centerline{\includegraphics[width=\columnwidth]{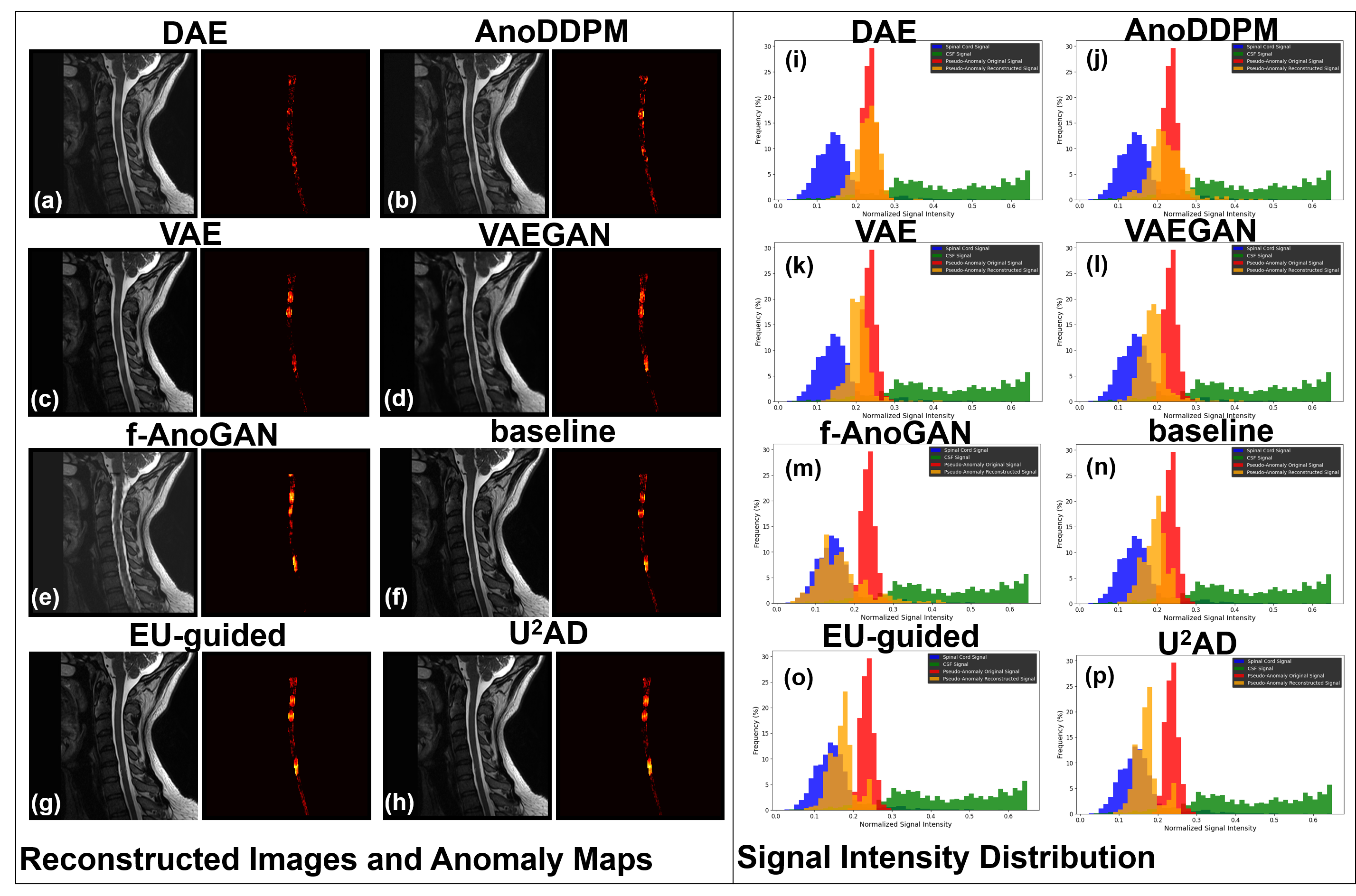}}
\caption{(a)-(h) Reconstruction results from different UAD models. The left column shows the reconstructed images, and the right column shows the corresponding anomaly maps. (i)-(p) Signal intensity distributions for the original (red) and reconstructed (orange) pseudo-anomaly regions.}
\label{openneuro_recon}
\end{figure*}

\subsection{Generation of pseudo-anomalies}\label{Generation_of_pseudo-anomalies}
The process of generating pseudo-anomalies involves creating both anomaly masks and anomalous signal distribution. The anomaly mask is elliptical in shape, with the width being a random value between 0.7 and 1 times the spinal cord width (w) at the placement location, and the length being a random value between 1 and 4 times the width. The pseudo-anomalous signal follows a Gaussian distribution. Initially, we compute the mean and variance of the signal distribution in the spinal cord and cerebrospinal fluid (CSF) regions. The T2 hyperintensity signal should have a higher intensity than the spinal cord but lower than the CSF, so the mean and variance of the generated pseudo-anomalous signal distribution are randomly selected between the values of the spinal cord and CSF. For each 2D image, we randomly generate 1 to 3 pseudo-anomalous regions and place them within the spinal cord region. The process is illustrated in the \Cref{pseudo-anomaly}. In the figure: (a) shows the original MR image. (b) provides the corresponding spinal cord (gray) and CSF (white) mask. (c) depicts a randomly generated pseudo-anomaly mask. (f) shows the generated pseudo-anomaly signal distribution. The pseudo-anomaly signals and masks are then combined and embedded into the MR image to generate pseudo-anomaly images, as shown in (c).

\subsection{Reconstruction analysis on pseudo-anomalies}\label{Reconstruction_analysis_on_pseudo-anomalies}
Using the example illustrated in \Cref{pseudo-anomaly}, we analyze the reconstruction performance of various UAD models specifically on pseudo-anomalous regions. To ensure a fair comparison, all models are trained following dataset training Strategy 3. The reconstruction results are shown in \Cref{openneuro_recon}. Among the tested models, DAE and AnoDDPM exhibit overfitting, as evident from their reconstructed results and signal distributions. VAE and VAEGAN demonstrate partial differentiation between normal SC signals and anomaly signals; however, the separation is not as pronounced as in f-AnoGAN and the proposed $U^{2}AD$ framework. f-AnoGAN achieves the closest reconstructed signal distribution to the normal SC signal, as shown in (m). However, its reconstructed image (e) appears blurry, potentially introducing noise into the anomaly maps and leading to misdiagnoses. In comparison, the proposed $U^{2}AD$ framework outperforms the baseline models. As shown in (h), $U^{2}AD$ achieves better reconstruction quality and anomaly detection performance. Its reconstructed pseudo-anomaly regions retain the signal intensity characteristics of normal SC regions, with anomaly maps effectively highlighting abnormal areas.

\section{Details of Comparison Methods}

\subsection{UAD methods:}\label{Details_of_UAD_methods}
For the denoising-based models DAE and AnoDDPM, we added noise to the entire image and obtained the denoised image from the noisy input. The ROI region in the denoised image was preserved, and the other areas were replaced with the original image, resulting in the final reconstructed image. For VAE and VAEGAN, we mapped the entire image to the latent space and generated the target image from the latent variables. Similarly, the ROI region of the generated image was retained, and the remaining part was replaced with the original image to obtain the reconstructed image. f-AnoGAN requires a two-step training process. The first step involves training a WGAN, where the input is a random vector \( z \), and the goal is to use the generator \( G \) to produce realistic ROI images that can deceive the discriminator \( D \). The second step trains an encoder to map the test image to the latent vector \( z \). We used the training strategy provided in the original paper (i.e., the IZI-F strategy), where we froze the parameters of the generator and discriminator, and only trained the encoder. We input the product of the original image and the ROI mask into the encoder, with the objective of making the generated latent vector \( z \) produce an image from \( G \) that closely matches the ROI region of the original image. Finally, the ROI region of the test image is input into f-AnoGAN to generate the ROI, and it is combined with the remaining areas of the original image to produce the reconstructed image. For AnoDDPM, we used 1000 diffusion steps (T=1000) for training, and during inference, we employed partial diffusion with 40 steps.

\subsection{Object detection methods:}\label{Details_of_Object_detection_methods}
Since our annotations are based on vertebral segments, the bounding boxes used for training correspond to the minimal enclosing rectangles of the spinal cord regions for the anomalous segments. We input the 2D MR images and corresponding bounding boxes into the models for training. During inference, the models output all candidate bounding boxes and their corresponding confidence scores for each test image. For patient-level anomaly identification, we use the highest confidence value from the object detection model as the anomaly score and determine whether the patient has an anomaly based on a chosen threshold. For segment-level anomaly localization, we use the bounding box with the highest confidence in the object detection results as the location of the anomaly. The predicted anomaly location must satisfy two conditions: first, the intersection-over-union (IoU) between the bounding box and the corresponding spinal cord region must exceed 0.5, and second, the confidence must be higher than the threshold. For the supervised object detection models, we also perform 5-fold cross-validation. The dataset is divided into 5 sub-folds, with 3 used for training, 1 for validation to select the optimal threshold, and the remaining fold used for testing.

\subsection{Signal analysis methods:}\label{Details_of_Signal_analysis_methods}
For the statistical approach, we selected Z-scoring, which is based on the standard normal distribution. It calculates the deviation of each pixel's intensity from the mean intensity of the ROI and normalizes it in terms of standard deviations to assess the anomaly level of the pixel. In this study, we used a standard deviation multiplier of 0.1 as the threshold to classify a pixel as anomalous. For clustering, we chose two methods: Simple Linear Iterative Clustering (SLIC) and Density-Based Spatial Clustering of Applications with Noise (DBSCAN). SLIC is a method that segments an image into superpixels based on the distance between pixels and signal intensity. We used the SLIC method to divide the ROI region in each image into \( M \) superpixel regions. For each superpixel region, we calculated its mean intensity and standard deviation, then used the Z-scoring method to evaluate the anomaly level of each region. In this study, we set \( M = 100 \) and the compactness parameter to 10, with a standard deviation threshold of 2 for Z-scoring. DBSCAN is a density-based clustering algorithm specifically designed to identify clusters and outliers (anomalies) in data. The two main hyperparameters of DBSCAN are \( \epsilon \) and \( min\_samples \), which control the size of the neighborhoods and the minimum number of points required to form a cluster. In this study, we set \( \epsilon = 0.01 \) and \( min\_samples = 100 \). For noise points, we calculated the distance to the nearest cluster as the anomaly score. Similar to the UAD methods, these approaches also produce anomaly maps and anomaly curves for the test images, which are then used for anomaly detection and localization.

\renewcommand{\thefigure}{A\arabic{figure}}
\begin{figure*}[!ht]
\centerline{\includegraphics[width=\columnwidth]{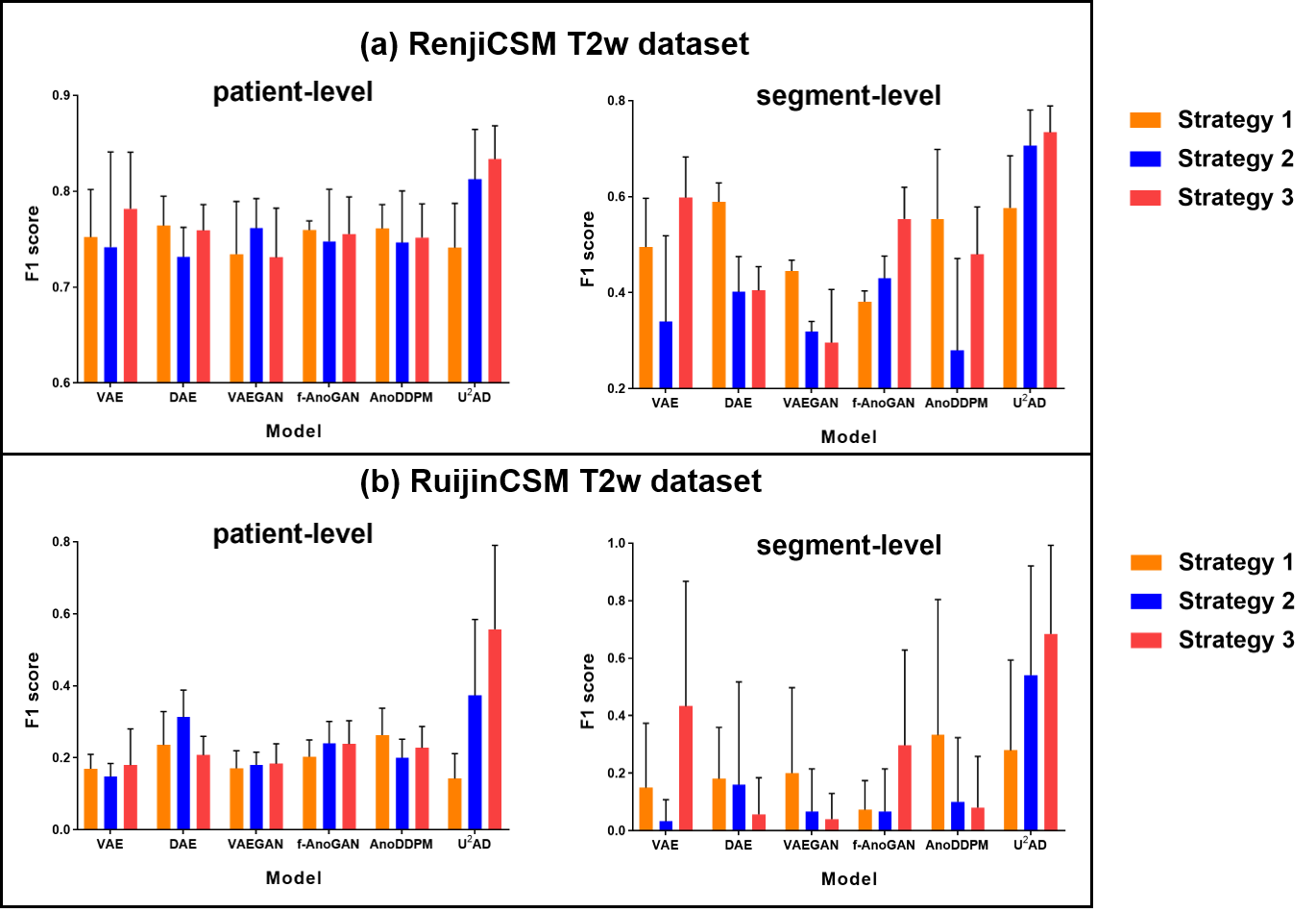}}
\caption{Comparison of different UAD methods under three dataset training strategies. Strategy 1 involves training models exclusively on the public healthy dataset (Spine Generic T2w). Strategy 2 involves training models only on the clinical dataset (RenjiDCM T2w). Strategy 3 combines pretraining on the public healthy dataset and adapting to the clinical dataset. All models are evaluated on the clinical datasets to assess their performance in patient-level identification and segment-level localization tasks.}
\label{performance_comparison}
\end{figure*}

\section{Comparison of three dataset training strategies}
\Cref{performance_comparison} further visualizes the impact of three dataset training strategies on the performance of the six UAD models. The best dataset training strategy for each UAD model varies. From the results of RenjiDCM T2w dataset, VAE and $U^{2}AD$ performed best with strategy 3, while DAE, f-AnoGAN, and AnoDDPM performed best with strategy 1. VAEGAN performed best using strategy 2. It is worth noting that $U^{2}AD$ did not achieve the best results in all three training strategies. When using strategy 1, $U^{2}AD$ achieved F1 scores of 0.7411 for patient-level and 0.5770 for segment-level on RenjiDCM T2w datset, which were lower than those of DAE and other models. However, after adapting on clinical data, $U^{2}AD$ showed significant improvements across all metrics.

\renewcommand{\thefigure}{A\arabic{figure}}
\begin{figure*}[!t]
\centerline{\includegraphics[width=\columnwidth]{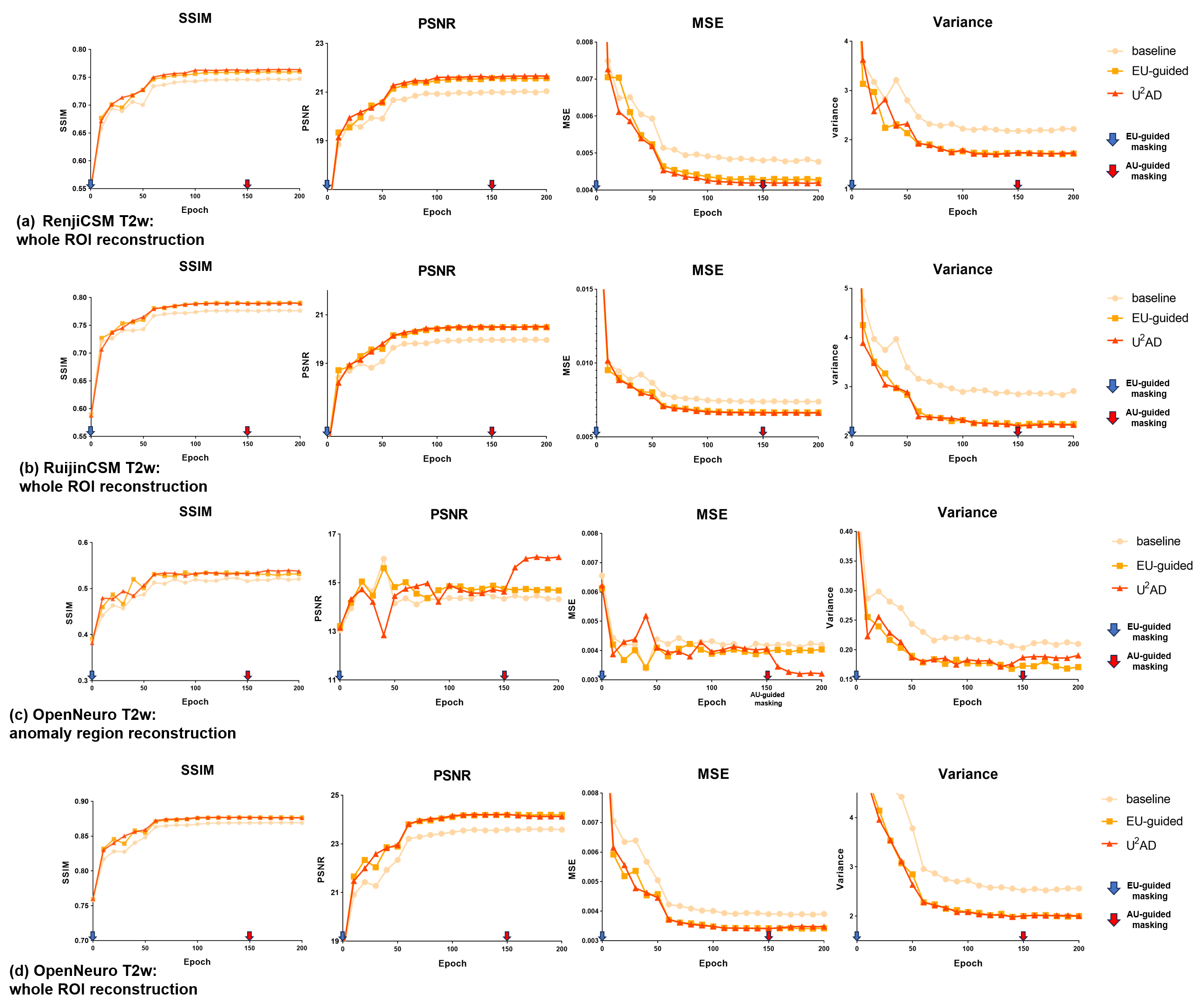}}
\caption{Assessment and comparison of reconstruction performance among the baseline, EU-guided, and $U^{2}AD$ models. (a) and (b) illustrate the reconstruction performance curves for the whole ROI on the RenjiDCM T2w dataset and Ruijin T2w dataset respectively, evaluated using metrics including SSIM, PSNR, MSE, and Variance. (c) and (d) present the reconstruction performance curves for anomaly regions and the whole ROI, respectively, on the OpenNeuro T2w dataset. The baseline model adopts a random masking strategy throughout the process, while both EU-guided and $U^{2}AD$ models utilize the EU-guided masking strategy from epoch 0. Additionally, $U^{2}AD$ incorporates the AU-guided masking strategy starting from epoch 150 to enhance anomaly detection performance.}
\label{reconstruction_trend}
\end{figure*}

\section{Effectiveness of Uncertainty-Guided Masking on Reconstruction}
\Cref{reconstruction_trend}(a) and (b) show the curves of reconstruction metrics and the total variance of the entire image during the adaptation on the RenjiDCM T2w and Ruijin T2w dataset. It can be seen that, in the clinical datasets, both EU-guided and $U^{2}AD$ outperform the baseline in SSIM and PSNR, while MSE and variance decrease. This indicates that the EU-guided masking improves the model's confidence in the reconstruction results (variance↓), leading to better reconstruction (SSIM↑, PSNR↑, MSE↓). \Cref{reconstruction_trend}(c) and (d) show the trend of reconstruction metrics in the OpenNeuro T2w dataset for pseudo-anomalous and normal regions, respectively. Specifically, in \Cref{reconstruction_trend}(c), the SSIM, PSNR, and MSE calculated on the pseudo-anomaly region between the reconstructed image and the original image (before adding the pseudo-anomaly) are shown. It shows that EU-guided outperforms the baseline in every metric. On the other hand, after the inclusion of AU-guided masking (at epoch 150), the PSNR for the pseudo-anomaly region increases significantly, and MSE drops sharply, indicating that AU-guided masking effectively improves the reconstruction of anomalous regions. \Cref{reconstruction_trend}(d) shows the quantitative results for the whole ROI reconstruction, where both EU-guided and $U^{2}AD$ improve the reconstruction metrics, indicating the reconstruction of normal regions is not affected by AU-guided masking after epoch 150.

\renewcommand{\thefigure}{A\arabic{figure}}
\begin{figure*}[ht]
\centerline{\includegraphics[width=\columnwidth]{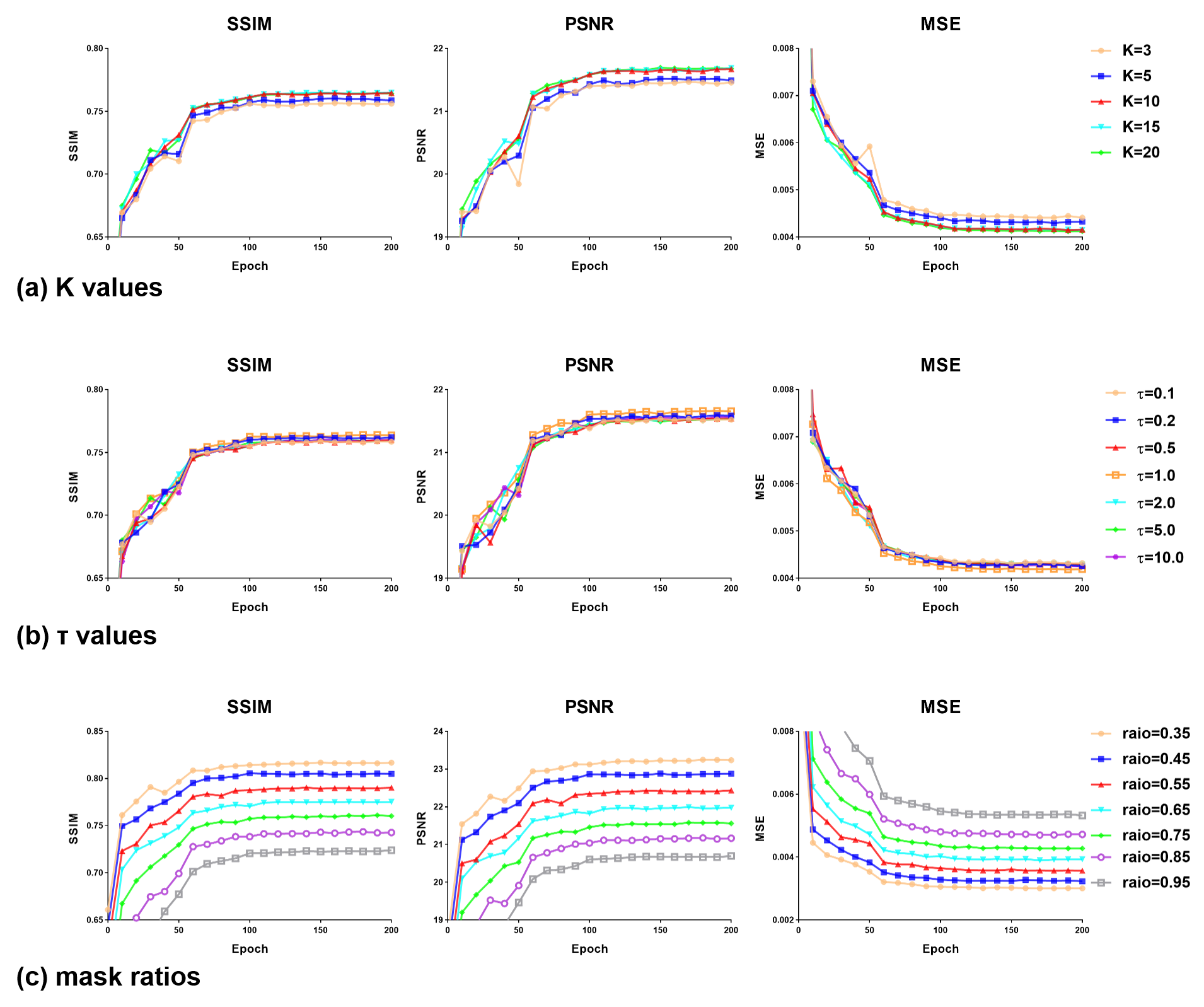}}
\caption{Comparison of reconstruction performance under varying hyperparameter settings, including \(K\) values (number of reconstructions), $\tau$ values (temperature factor in EU-guided masking), and mask ratios. (a) illustrates the impact of different \(K\) values on reconstruction metrics, highlighting the trade-off between computational cost and model performance. (b) presents the effect of varying $\tau$ values on reconstruction performance, demonstrating its role in controlling the sensitivity of uncertainty estimation. (c) evaluates the influence of mask ratios on reconstruction fidelity and anomaly detection accuracy, showing how different levels of masking affect model robustness.}
\label{hyperparameters}
\end{figure*}

\section{Comparison of Reconstruction Performance of Different Hyperparameters}
In \Cref{hyperparameters} (a), we observe that the reconstruction results of our model are not sensitive to \(K\). When \(K\) increases from 3 to 5 and then to 10, SSIM and PSNR show slight improvements, while MSE decreases slightly. However, no further improvements are observed when $K$ is increased beyond 10 (i.e., for $K = 10$, $K = 15$, and $K = 20$). In \Cref{hyperparameters} (b), we show the impact of different $\tau$ values on the model's reconstruction results. It can be seen that $\tau$ does not significantly affect reconstruction performance in terms of SSIM, RSNR, and MSE. However, it is worth noting that when $\tau = 0$, the probability weights of different patches are uniform, effectively becoming random masking as in the baseline model. From \Cref{reconstruction_trend}, we can observe that EU-guided and $U^{2}AD$ with $\tau > 0$ show significant improvements in reconstruction results compared to the baseline with $\tau = 0$. \Cref{hyperparameters} (c) demonstrates that higher mask ratio leads to worse reconstruction within the masked regions.

\renewcommand{\thefigure}{A\arabic{figure}}
\begin{figure*}[!ht]
\centerline{\includegraphics[width=0.7\columnwidth]{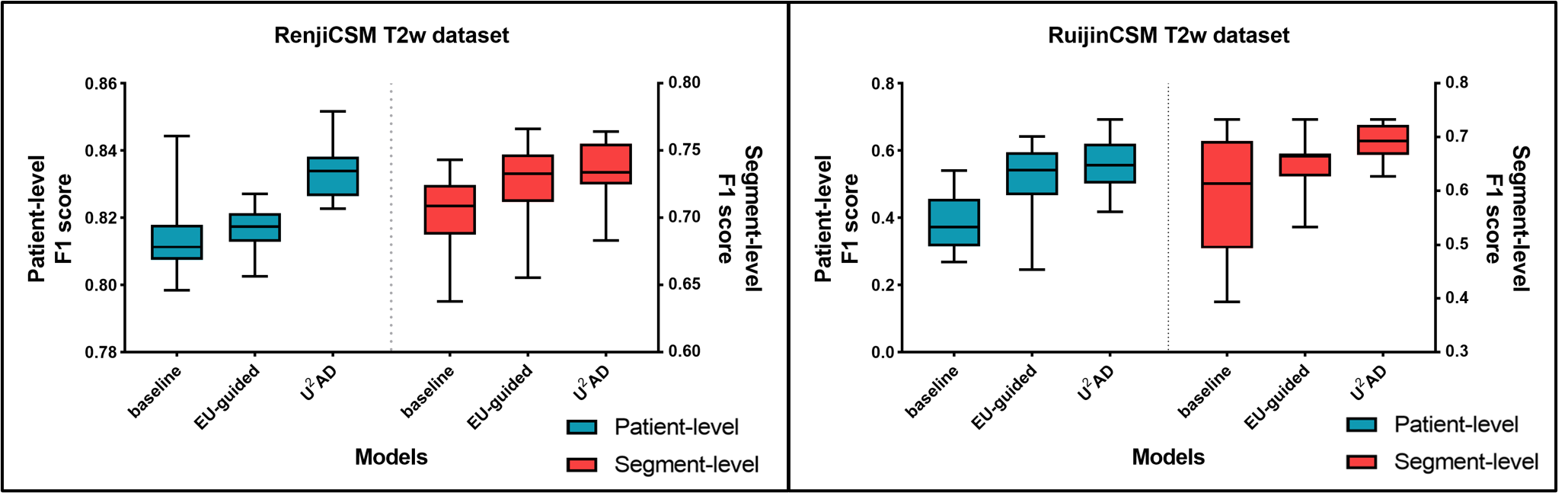}}
\caption{Stability assessment of the trained model through 20 repeated iterations of five-fold cross-validation. The box plots illustrate the distribution of the 20 mean values for the patient-level F1 score (blue) and segment-level F1 score (red). The results highlight the robustness and consistency of the model, showing minimal variability in both metrics across repeated iterations.}
\label{boxplot}
\end{figure*}

\section{Model Reliability}
Due to the randomness of Monte-Carlo inference, the reconstruction results and anomaly maps can vary, leading to some variation in experimental outcomes. To assess the stability of our model in the anomaly detection task, we performed 20 repeated iterations of five-fold cross-validation with the trained model, and all results are presented in \Cref{boxplot}. Among the three models compared, $U^{2}AD$ demonstrated the highest reliability. On the RenjiDCM T2w dataset, even the worst result of $U^{2}AD$ had an patient-level F1 score greater than 0.82, demonstrating the stability of our model in the patient-level anomaly detection task. In the segment-level localization task, the model's stability showed a slight decline. The lower quartile F1 score of $U^{2}AD$ was around 0.72, and the worst result had an F1 score greater than 0.68. On the RuijinDCM T2w dataset, where the anomaly prevalence is sparse, the stability of baseline and EU-guided models declined significantly due to the challenges posed by the dataset's low anomaly density. In contrast, $U^{2}AD$ remained consistent across repeated assessments, achieving stable and reliable results in both patient-level identification and segment-level localization tasks.

\bibliographystyle{elsarticle-harv}
\bibliography{sample}

\end{document}